\documentclass[12pt,aps,onecolumn,pra,superscriptaddress,nofootinbib,floats,notitlepage,preprintnumbers]{revtex4-1}
\usepackage{amsmath,amssymb,color,mathrsfs, graphicx,verbatim,epsfig, bbm, wasysym,simplewick}
\usepackage{slashed}
\usepackage{fbb}
\allowdisplaybreaks

\newcommand{\vev}[1]{ \left\langle {#1} \right\rangle }
\newcommand{\bra}[1]{ \langle {#1} | }
\newcommand{\ket}[1]{ | {#1} \rangle }
\newcommand{\be}{\begin{eqnarray}}
\newcommand{\ee}{\end{eqnarray}}
\newcommand{\nn}{\nonumber}
\newcommand{\kir}{k_{\rm IR}}
\newcommand{\pmv}{P \! - \! V}
\newcommand{\tor}{\rightarrow}
\def\beq{\begin{equation}}
\def\eeq{\end{equation}}
\def\bit{\begin{itemize}}
\def\eit{\end{itemize}}
\newcommand{\mo}{\mathcal{O}}

\begin{document}
\preprint{UMD-PP-019-10}

\title{De Sitter Diagrammar and the Resummation of Time}

\author{Matthew Baumgart}
\affiliation{Department of Physics, Arizona State University, Tempe, AZ 85287}
\author{Raman Sundrum}
\affiliation{Maryland Center for Fundamental Physics, University of Maryland, College Park, MD 20742} 

\begin{abstract}

Light scalars in inflationary spacetimes suffer from logarithmic infrared divergences at every order in perturbation theory.  This corresponds to the scalar field values in different Hubble patches undergoing a random walk of quantum fluctuations, leading to a simple toy ``landscape'' on superhorizon scales, in which we can explore questions relevant to eternal inflation.  However, for a sufficiently long period of inflation, the infrared divergences appear to spoil computability.   Some form of renormalization group approach is thus motivated to resum the log divergences of conformal time.  Such a resummation may provide insight into De Sitter holography.  We present here a novel diagrammatic analysis of these infrared divergences and their resummation.  Basic graph theory observations and momentum power counting for the in-in propagators allow a simple and insightful determination of the leading-log contributions.  One thus sees diagrammatically how the superhorizon sector consists of a semiclassical theory with quantum noise evolved by a first-order, interacting classical equation of motion.  This rigorously leads to the ``Stochastic Inflation'' ansatz developed by Starobinsky to cure the scalar infrared pathology nonperturbatively.  Our approach is a controlled approximation of the underlying quantum field theory and is systematically improvable.  
\end{abstract}

\vspace*{0.05in}
\maketitle

\section{Introduction}
\label{sec:intro}

De Sitter may be thought of as the spacetime with the best sense of humor.  A beautifully, indeed {\it maximally}, symmetric solution to Einstein's equations, it nonetheless gives an excellent approximate description to both our cosmological past, through inflation, and our future, by the coming era of dark energy domination.  However, at the quantum level, it poses a challenging set of questions, even in the infrared, where a UV-complete theory of quantum gravity seems unnecessary.  All observers see a horizon, which leads to questions about the precise nature of De Sitter (DS) temperature and the appropriate microstate description.  These horizons, along with the spacelike boundaries of the global spacetime, have made the proper holographic description elusive.  Even at the level of perturbatively computing correlation functions, certain quantum field theories in DS face large infrared sensitivities that grow with time.  The purpose of this work is to understand how one properly computes in one class of these theories, very light, non-derivatively-interacting scalars on a fixed De Sitter background.\footnote{As shown in \cite{Urakawa:2009my,Giddings:2010nc,Urakawa:2010kr,Senatore:2012nq,Pimentel:2012tw,Senatore:2012ya,Assassi:2012et,Tanaka:2013xe,Tanaka:2017nff}, single-field inflation does not suffer from large infrared sensitivities in perturbation theory because the inflaton is determining the geometry.  However, this perturbative breakdown would arise in an inflationary theory with a light spectator scalar for a sufficiently long period of inflation.}  The main results are technical, and yet provide suggestive hints for some of the deeper conceptual issues in quantum De Sitter correlators and cosmology.

Before describing the connections to important topics such as eternal inflation, the measure problem, and holography, we state that our resolution to the infrared pathologies of certain De Sitter theories follows the familiar formulation of ``stochastic inflation,'' developed originally by Starobinsky in the mid-1980s \cite{Starobinsky:1986fx}, and further elucidated by \cite{Starobinsky:1994bd,Tsamis:2005hd}. For an overview of more recent literature, see \cite{Hu:2018nxy}.  What is novel in our work is a rigorous, all-orders, diagrammatic derivation of the evolution equation for light-scalar correlation functions in De Sitter.  A key simplifying feature in our presentation comes from the constraints of manifest causality \cite{Musso:2006pt}, following from the reorganization of in-in perturbation theory given by Weinberg \cite{Weinberg:2005vy}.  Another key feature is identifying the simple structure of propagators and vertices in the soft limit, with a careful accounting of where the soft approximation breaks down within hard loops ({\it cf.}~Appendix \ref{app:closel}).   Recast as the evolution of a generating function, at leading-order in controlled approximations that we make explicit, we recover Starobinsky's Fokker-Planck equation for stochastic inflation.  An earlier diagrammatic approach to deriving stochastic inflation can be found in \cite{Garbrecht:2013coa,Garbrecht:2014dca}, but without manifest causality it has a different character.  As we will show, the leading infrared contributions to correlation functions are given by the convolution of causal, classical perturbative evolution with quantum noise.  Using an a non-manifestly-causal basis, this simple property is obscured in these earlier diagrammatic papers.  Furthermore, the graphical analysis in them maximizes the IR enhancement at each vertex, whereas in our formulation, we find in Section \ref{sec:llcorr} that one must analyze a diagram globally to capture the dominant soft physics. 

\begin{figure}[ht!]
\begin{center}
\includegraphics[width=12cm]{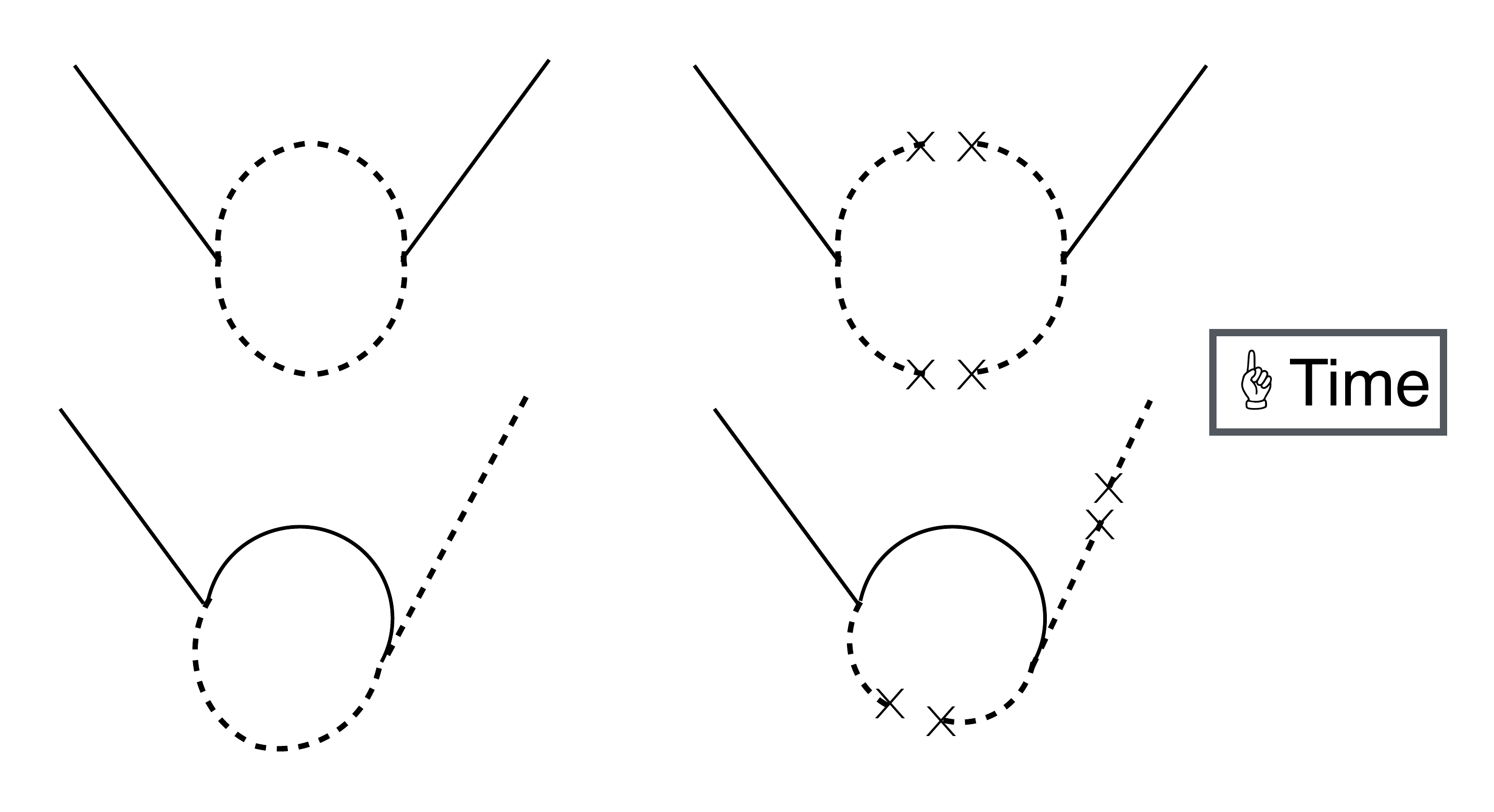}    
\end{center}
\caption{{\bf Left:} QFT Diagrams for $\vev{\phi(\eta, \vec x) \phi(\eta, \vec y)}$ in $\phi^3$ in-in perturbation theory, where time is flowing upward in the graph.  As explained in Section \ref{sec:setup}, propagators are either retarded, $G_R$ (solid lines), or anticommutator, $G_+ = \vev{\{ \phi_1,\, \phi_2 \}}$ (dashed lines).  {\bf Right:} As we will show in Section \ref{sec:llcorr}, in all leading contributions to $\vev{\phi^n}$ correlation function, the $G_R$ form tree shaped subdiagrams that touch one and only one correlation point.  Thus, if we cut each $G_+$ propagator and consider each ``$\times$'' as an insertion of the zeroth-order solution, $\phi^0$, we see that each diagram decomposes into a product of classical perturbation theory diagrams.  In Section \ref{sec:fp}, we detail how the inclusion of these $\phi^0$ as a quantum distribution leads to the famous Fokker-Planck description of De Sitter light scalar evolution.  The use of $\phi^3$ vertices is purely for visual simplicity.  The graph theory statements in this paper hold for a generic, non-derivatively coupled scalar potential.  Although a $\phi^3$ interaction is unstable, one can nonetheless consider it as a subsector of an ultimately (meta)stable theory.
\vspace{0.025in}}
\label{fig:scqft}
\end{figure}

The key physical insight of stochastic inflation is that superhorizon modes in De Sitter follow a first-order, inhomogeneous classical equation of motion.  The inhomogeneity is given by a stochastic source with a known distribution.  Its intrinsic randomness is the remaining quantum feature in the problem.  It reflects the fact that all comoving modes in De Sitter redshift, and even those in the UV that have heretofore admitted a healthy perturbative description will eventually ``fall'' into the nonperturbative regime.  This entry into the strongly-interacting superhorizon sector can still be described by perturbation theory though.
By power counting, we will derive the structure of De Sitter Feynman diagrams that has the leading sensitivity to the infrared breakdown of perturbation theory (or equivalently, the leading secular growth).  The ingredients for this are nothing other than causality and the momentum scaling of two different types of propagators that arise in the in-in formalism for correlators.  These leading diagrams then make sharp the sense in which the infrared of De Sitter field theory has a predominantly semiclassical description.  We will see that an arbitrary diagram with an arbitrary number of loops contains within it a substructure of tree diagrams with all non-tree features simply corresponding to the inhomogeneous stochastic noise sources that arise from the redshifting of UV modes ({\it cf.}~Fig.~\ref{fig:scqft}).  
As mentioned above, this identification of the leading behavior allows one to compute the evolution equation for any coincident $\vev{\phi^n}$ correlation function, and ultimately that of a generating function, $p(\phi,t)$ such that $\vev{\phi^n} = \int d\phi \, p(\phi,t) \phi^n$.  This Fokker-Planck evolution for $p(\phi,t)$ can be solved at late times, showing that De Sitter correlation functions remain bounded, well-behaved and De Sitter invariant.

The ability to predict a distribution in field-strength $\phi$ at late times has important implications for Eternal Inflation\cite{Creminelli:2008es,Dubovsky:2008rf,Dubovsky:2011uy}.\footnote{For an overview of issues related to Eternal Inflation, see \cite{Guth:2007ng} and references therein.}  In particular, for the toy model of a massless De Sitter scalar we have a solved example of a landscape and a resolution to its measure problem.  Our generating function, $p(\phi, t)$ is the wavefunctional mod squared for superhorizon modes, meaning that in position space it provides a distribution over Hubble patch averages of field strength.  Furthermore, this gives the different probabilities on the future boundary of the spacetime in the limit $t \to \infty$.  To make contact with a more realistic measure problem, one could relax the assumption of a fixed background geometry with constant Hubble parameter, $H$.  By  accounting for the backreaction of the field-strength on the geometry, one is immediately dealing with a distribution over Hubble patches of different energy densities and curvature.  While there may be significant technical refinements needed for practical computation, we have reduced questions about the landscape and the ``wavefunction of the universe'' to perturbations, which one can hope are  IR-resummable, in a manner building on our analysis of fixed-background scalar dynamics and the well-defined framework of stochastic inflation.       

Having the wavefunctional mod squared on the boundary of De Sitter has important implications for holography as well.  A detailed understanding of the holographic dual of De Sitter is still an open question.  A particular challenge arises from the fact that boundary is spacelike, which suggests that time is holographically emergent the De Sitter bulk.  In its most limited sense, stochastic inflation provides a way to compute valuable theoretical data in the form of these probability distributions.  However, there is already a tantalizing hint about the dual-theory dynamics.  Stochastic inflation is Markovian.  Our rigorous approach to capture the leading behavior in correlation functions leads us to drop quantum phase information and work directly in terms of probabilities.  This is reminiscent of the parton shower of QCD, where the time evolution follows a first-order differential equation.  The connection between Stochastic Inflation and the parton shower has been noted before \cite{Seery:2009hs}, but implications for holography are still to be elucidated.

As a guide for the reader, in Section \ref{sec:setup} we show how an infrared divergence appears even at the level of free field theory of a massless scalar in De Sitter.  Section \ref{sec:llcorr} establishes how the leading infrared sensitivity for correlation functions of interacting light scalars in De Sitter follows from causality and power counting.  Then in Section \ref{sec:scfo}, we show how the physics that gives the leading-log behavior is semiclassical and described by a first-order equation of motion.  Putting all these properties together in Section \ref{sec:fp}, we derive the time-evolution of arbitrary, coincident, correlation functions, which we extend to the evolution of the generating function, $p(\phi,t)$.  We thus recover the central result of stochastic inflation, but from a fully first-principles QFT calculation.  Lastly in Section \ref{sec:conc}, we physically interpret our technical results and discuss future directions.

As this work was in final preparation, a related work was released \cite{Gorbenko:2019rza}.   Their emphasis is different, adopting a wavefunctional/path integral approach to understanding stochastic inflation, building on \cite{Riotto:2011sf,Moss:2016uix,Tokuda:2017fdh}.  In particular, they do not adopt a fully diagrammatic framework, and discussions of semiclassicality rely on statements made at fixed order in perturbation theory, rather than our all-orders approach.  However, they have also obtained some further generalizations and corrections, such as the ability to compute correlation functions at noncoincident times.

\section{Massless Scalar in De Sitter Spacetime}
\label{sec:setup}

We first define some notation, taking the FRW (or ``planar'') coordinates for De Sitter (DS) space,  
\be
ds^2 &=& dt^2 - e^{2Ht}d{\bf x}^2 \nn \\
 &=& \frac{1}{(H \eta)^2} [d\eta^2 - d{\bf x}^2],
\ee
where we frequently utilize conformal time,
\begin{equation}
 \eta \equiv -\frac{e^{-Ht}}{H}.
\end{equation}
In real-time, but spatial-momentum space, the action for our theory is thus,
\beq
S \,=\, \int_{\frac{1}{k_{\rm IR}}} \!\!\!\! d\eta \, \int_{k_{\rm IR}} \! \frac{d^3 k}{(2\pi)^3} \, \frac{1}{(H \eta)^2} \frac{1}{2} \Big [ (\partial_\eta \phi)^2 - k^2 \phi^2 \Big ] - \frac{1}{(H\, \eta)^4}V(\phi) 
\label{eq:dsaction}
\eeq
where we have imposed a comoving IR regulator, not only spatially, but also temporally in the form of an earliest time.   Its full significance and physical interpretation will be explained below. It explicitly breaks DS isometries of course, but these are recovered if we can ultimately take $\kir \rightarrow 0$.

From the free action of a massless scalar,\footnote{All equations in this paper are for the strictly massless case.  Nonetheless, the same perturbative breakdown that infects the $m=0$ theory occurs if $m \ll H$, as well \cite{Starobinsky:1994bd,Burgess:2010dd}.} we obtain the equation of motion for $\phi$, 
\be
\partial_t^2 \phi + 3H \partial_t \phi + e^{-2Ht} k^2 \phi &=& 0 \nn \\
\partial_\eta^2 \phi - \frac{2}{\eta} \partial_\eta \phi + k^2 \phi &=& 0.
\label{eq:eom}
\ee
This has the independent solutions
\be
\phi_1 &=& \frac{1}{\sqrt{2}} H \eta^{3/2} [-i (k \eta)^{-3/2} - (k \eta)^{-1/2}] e^{i k \eta}, \nn \\
\phi_2 &=& \frac{1}{\sqrt{2}} H \eta^{3/2} [i (k \eta)^{-3/2} - (k \eta)^{-1/2}] e^{-i k \eta},
\label{eq:dsmodes}
\ee
corresponding to positive and negative frequencies in the far past as the mode wavelengths blueshift to be well inside the horizon.
Canonical quantization is then defined by the linear combination,
\be
\phi_k = \phi_2 \, a_k + \phi_1 \, a^\dag_{-k},
\ee
where the coefficients are destruction/creation operators, $\left[ a_k,\, a^\dag_p \right] = (2\pi)^3 \delta(k-p)$.  The state annihilated by all the $a_k$ defines the so-called ``Bunch-Davies vacuum'' \cite{Bunch:1978yq}. As wavelengths blueshift well inside the horizon in the arbitrary past, this vacuum asymptotes to that of Minkowski space.

We are now in a position to compute the basic Wightman function,
\beq
G_W(k;\, \eta^\prime, \eta) \equiv \vev{{\rm BD}| \phi_k(\eta^\prime) \phi_{-k}(\eta)|{\rm BD}} = \frac{H^2}{2 k^3} e^{-ik(\eta^\prime - \eta)} \Big[ 1+ik \, (\eta^\prime - \eta)+ k^2 \, \eta^\prime \eta \Big ].
\label{eq:momwight}
\eeq
In combination with the time-ordering and $\theta$ functions, one can construct any two-point function from it.  In this paper, we will find it most useful to work in terms of the following anti-symmetric and symmetric combinations of this, to define
\be
G_R(k; \, \eta^\prime, \eta) &\equiv& \theta(\eta^\prime - \eta) \, [\phi_k(\eta^\prime), \phi_{-k}(\eta)] \nn \\
G_+(k; \, \eta^\prime, \eta) &\equiv& \vev{\{ \phi_k(\eta^\prime), \phi_{-k}(\eta) \}},
\ee
where we note that the retarded propagator, $G_R$, is independent of the external state.
 
The pathology of a massless scalar in DS is apparent already at the free two-point level if we do not IR regulate the theory.  Fourier transforming the Wightman function leads to an IR divergence,
\be
G_W(\eta^\prime, x^\prime;\, \eta, x) &=& \int \frac{d^3k}{(2\pi)^3} \, e^{ik(x^\prime-x)} \, G_W(k;\, \eta^\prime, \eta) \nn \\
&\underset{k \rightarrow 0}{\sim} & \int \frac{dk}{2 \pi^2} \frac{H^2}{2k}.
\label{eq:poswight}
\ee
We see that this is logarithmically infrared divergent, but can be rendered finite with a comoving cutoff, $\kir$.
Upon adding interactions, we will find that any fixed order of perturbation theory will contain sensitivity to $\kir$ of the form $\log(\kir)^n$.  However, we will show that after resumming the leading-log contributions to all orders, we obtain a finite expression for $\kir \tor 0$, for nonpathological potentials.  Thus, one can regard $\kir$ as a formal regulator needed at intermediate steps, but removable in a complete calculation.\footnote{Despite the appearance of $\kir$ in the free theory, due to the shift symmetry of a free, massless scalar, $\kir$ can also be removed in physical quantities related to the energy-momentum tensor.}  

It is illuminating though, to consider the possible physical origins and interpretation of such a cutoff, and why it should regulate comoving rather than physical momenta.  Instead of taking a pure De Sitter spacetime, we can posit an earlier epoch of non-DS, IR-safe geometry that transitions to DS at a ``start of inflation,'' $\eta_0$.  We can then ask how sensitive late-time (meta)observables are to the details of this cosmic transition to DS.  Operationally, this pre-DS era provides an IR regulation of the integral in Eq.~\ref{eq:poswight}.  For example, this Universe could have been in a radiation-FRW phase early on ($\rho_{\rm tot} = \rho_{\rm rad} + \rho_\Lambda$), but after sufficient redshifting of the radiation, the cosmic evolution would be dominated by the cosmological constant, and would thus enter a DS expansion.\footnote{One can alternatively consider a Universe born at a finite time with a wavefunctional giving an IR-safe, but DS-like-in-the-UV theory.  At the free-theory level, we would have $\Psi[\{\phi_k\},t] \propto \exp[\int d^3k f(k,\,t) \phi_k \phi_{-k}]$, where the function $f(k,\, t)$ recovers the Bunch-Davies Wightman function (Eq.~\ref{eq:momwight}) for $k \gg 1/\eta_0$, but diverges less severely than $1/k^3$ at small $k$.}  Since this theory (like pure radiation FRW) is IR-safe, we can find some $\kir$ such that simply cutting off modes with $k<\kir$ introduces only small power corrections, $\kir^n$, 
relative to a complete matching procedure between pre-inflationary and inflation phases.  
To choose the value of $\kir$, we note that the modes already outside the horizon at the start of inflation, $k \eta_0 \ll 1$, never re-enter the horizon during the DS era and remain frozen ({\it cf.}~Eq.~\ref{eq:dsmodes}) in their IR-safe configuration. These are just the modes that we can safely drop up to small power corrections. That is we should take, 
\begin{equation}
\kir \eta_0 \sim {\cal O}(1), 
\end{equation}
giving a physical interpretation of the regularization of the action in Eq.~\ref{eq:dsaction}.

Thus, we can write a regulated Wightman function,  
\beq
G_W(k;\, \eta^\prime, \eta) \equiv \vev{\phi_k(\eta^\prime) \phi_{-k}(\eta)} = \frac{H^2 \, \theta(k - \kir)}{2 k^3} e^{-ik(\eta^\prime - \eta)} \Big[ 1+ik \, (\eta^\prime - \eta)+ k^2 \, \eta^\prime \eta \Big ], 
\label{eq:momwightreg}
\eeq
where we assume that $\eta, \eta'$ are in the DS era, later than $\eta_0$. 
This matches a DS-transitioning cosmology as sketched above, up to ${\cal O}(\kir^{n > 0})$ corrections.  Since we are tracking and resumming logarithmic dependence on $\kir$ in this paper, such power corrections are negligible. 
This equation defines the modified state in which we will compute expectation values, which we denote $|{\rm BD}^\prime \rangle$. It regulates the IR compared to the standard Bunch-Davies vacuum, whose Wightman function is given in Eq.~\ref{eq:momwight}.  Except where noted, we will drop the explicit $|{\rm BD}^\prime \rangle$ and just evaluate expectations with $\langle ...\rangle$.
Upon Fourier transforming, we now get a finite result, but one that is logarithmically sensitive to our cutoff,
\be
G_W(\eta^\prime, x; \, \eta, x^\prime) &=& \frac{-H^2}{4\pi^2} \left( \frac{\eta^\prime \, \eta}{(\eta^\prime - \eta -i\epsilon)^2 - (x^\prime -x)^2} \right. \nn \\
&&\left. + \frac{1}{2} \log \big[ \kir^2 \big( (\eta^\prime - \eta -i\epsilon)^2 - (x^\prime -x)^2 \big) \big] \right),
\label{eq:wightschem}
\ee
where we again neglect any arising ${\cal O}(\kir^{n > 0})$ pieces. 

Once we include interactions, we will find new IR divergences beyond those of the free theory, but they remain logarithmic. 
The focus of this work will be to track the leading sensitivity to $\kir$ in correlation functions. For example, in $\lambda \phi^4$ theory we will show that such terms are of the form $\lambda^r \log(\kir \eta)^s$ in (a class of) expectation values at time $\eta$ well into the DS era, $\eta/\eta_0 \sim \eta k_{IR} \ll 1$. That is, we are equivalently tracking large $(\log(\eta/\eta_0))^s \sim (t - t_0)^s$. We will demonstrate that in the late-time limit, the leading-log contributions are resummable and one obtains a finite result for $\eta \rightarrow 0$.\footnote{There is an important caveat here as this analysis holds for leading-logs only.  There is a possibility that {\it subleading} logs still give a late-time ($\eta \rightarrow 0$) divergence.  We leave this question for future work.  At minimum, what we have shown is that the theory is trustworthy up until the next-to-leading-log (NLL) corrections become important.}  This is identical mathematically to taking $\kir \tor 0$, removing the cutoff, pushing the pre-DS cosmology infinitely far into the past.  Thus, whether we are interested in a pure, infinite-time DS, or an inflation-like scenario with a finite duration, we can adopt the same approach so long as inflation lasts for sufficient time to make the leading-log analysis relevant.

\section{Leading-Logs to All Orders}
\label{sec:llcorr}

A natural set of observables for a DS phase of cosmology are in-in correlators ({\it i.e.}~computed with the same quantum state in both bra and ket)
of products of local operators in the late-time limit. 
However, performing an experiment to test this regime requires the De Sitter era to ultimately end, as in the inflationary paradigm, after which the correlations can re-enter our single horizon. We do not explicitly alter the future geometry to study this latter process in detail.  Nonetheless, we assume it can 
occur in principle, making the DS metaobservables, $\langle \phi(\eta,x_1) ... \phi(\eta,x_N)  \rangle$, legitimate to study at some late time $\eta$. 
For simplicity (and yet already subtle) we focus on observables  $\langle \phi(\eta,x)^n  \rangle$, where all operators are coincident in both time and space.

We can state our central result for these correlation functions up front, which we will prove to all orders in perturbation theory:
\begin{quote}
{\it Theorem: For any perturbative in-in diagram that contributes to $\vev{\phi^n}$ at a coincident space-time point, its maximal sensitivity to the comoving infrared cutoff, $\kir$, is $\log(\kir)^P$, where $P$ is the number of propagators in the graph.  Furthermore, these ``leading-log'' graphs contain exactly $V$ retarded propagators, where $V$ is the number of vertices, with at least one retarded propagator touching every vertex and some external correlation point (i.e. one of the $\phi(\eta,x)$ in the observable, $\phi^n$).  The retarded propagators form tree subdiagrams, with each tree touching one and only one correlation point.  The trees are joined together to make a complete diagram by anticommutator two-point functions, $G_+ = \vev{\{ \phi_1,\, \phi_2 \}}$.}
\end{quote}
As is familiar from general perturbative classical field theory, the appearance of trees of retarded propagators is the perturbative face of nonlinear {\it classical} evolution. But the dressing of these retarded trees by $G_+$ propagators reflects that the nonlinear classical evolution is being seeded by a non-classical, quantum, source. We will show how the classical and quantum features combine as part of the leading-log resummation.

\subsection{In-In Perturbation Theory}
\label{subsec:iigt}

To begin, we start with a theory we can treat perturbatively for sufficiently small coupling.  Formally, the $\phi^n$ expectation value within the {\it interacting} ``vacuum,'' $\ket{\Omega}$, is given in the interaction picture by
\be
\bra{\Omega} \phi_{\rm Heis.}(t,x)^n  \ket{\Omega} &=& \lim_{t_0 \to -\infty} \bra{\rm BD} \left[ \bar T \exp \left( i \int_{t_0 (1+i\epsilon)}^t H_I(t^\prime) dt^\prime \right) \right] \nn \\
&&\times \phi_I(t,x)^n \left[ T \exp \left( -i \int_{t_0 (1-i\epsilon)}^t H_I(t^{\prime\prime}) dt^{\prime\prime} \right) \right] 
\ket{\rm BD}/ \, \mathcal{N}, 
\label{eq:preintev}
\ee
where $\phi_I(t,x)$ is only evolved by the free Hamiltonian, and where $T,\, \bar T$ are time and anti-time ordering.  We have written this in terms of the more canonical proper time $t$, for familiarity's sake.  The $\epsilon$ factor in Eq.~\ref{eq:preintev} plays the usual role of projecting free Bunch-Davies state onto the interacting vacuum, and $\mathcal{N}$ is the usual division by (and hence dropping of) vacuum bubble contributions.  We can clearly expand the correlator perturbatively in the interaction Hamiltonian, $H_I$, to any fixed order.  

However, the resulting contributions will suffer IR divergences, and thus require regulation, as we elaborate below.  We will therefore replace Eq.~\ref{eq:intev} with 
\be
\bra{\Omega} \phi_{\rm Heis.}(t,x)^n  \ket{\Omega}\Big |_{\rm IR-reg.} &=& \bra{\rm BD^\prime} \left[ \bar T \exp \left( i \int_{t_0 (1+i\epsilon)}^t H_I(t^\prime) dt^\prime \right) \right] \nn \\
&&\times \phi_I(t,x)^n \left[ T \exp \left( -i \int_{t_0 (1-i\epsilon)}^t H_I(t^{\prime\prime}) dt^{\prime\prime} \right) \right] 
\ket{\rm BD^\prime} /\mathcal{N}, 
\label{eq:intev}
\ee
where $|{\rm BD}^\prime \rangle$ is the state defined by Eq.~\ref{eq:momwightreg}, the prime instructing us to cut off comoving momenta below $\kir$.  As noted in the discussion above Eq.~\ref{eq:momwightreg}, the regulated initial time, $t_0$ is related to the comoving momentum cutoff, $\kir$, by $t_0 \sim -1/H \log(H/\kir)$.  

In any logarithm, the dimensions of $\kir$ are only balanced in a coincident expectation value, $\vev{\phi_{\rm Heis.}(t,x)^n}$, by $\eta$, the only available scale. To see this, note that  
by spatial translation invariance, there can be no dependence on $x$.  De Sitter does contain an intrinsic dimensionful scale, $H$, but we see it only enters the Wightman function (Eq.~\ref{eq:wightschem}) as an overall factor, and can thus be scaled away and put back by dimensional analysis at the end.  We will see that the breakdown in perturbation theory is no worse than logarithmic, {\it i.e.}~there will be no power-law dependence on $\kir$.

The causal structure underlying expectation values in the above canonical form is more straightforwardly seen by a re-expression in terms
a series of nested commutators \cite{Weinberg:2005vy},
\be
\vev{\phi_{\rm Heis.}(t,x)^n} &=& \sum_{V=0}^\infty (-i)^V \int_{t_0}^t dt_V \ldots \int_{t_0}^{t_3} dt_2 \int_{t_0}^{t_2} dt_1 \nn \\
&\times & \Big \langle \Big [ \Big [ \ldots \Big [ \phi_I(t,x)^n,  H^\epsilon_I(t_V) \Big ] \ldots , H^\epsilon_I(t_2) \Big ], H^\epsilon_I(t_1) \Big ] \Big \rangle \nn \\
&\equiv & \sum_{V=0}^\infty \vev{\phi(t,x)^n}\big |_{\lambda^V},
\label{eq:pertex}
\ee
where the meaning of the $\epsilon$-deformation, $H^\epsilon_I$, revolves around the following important technical complication. Formally, in the absence of $\epsilon$, 
we demonstrate the equality of Eqs.~\ref{eq:intev} and \ref{eq:pertex} in Appendix \ref{app:inin}, by perturbative induction. However, this proof depends crucially on the 
 unitarity of the time evolution operator, which does not hold after the $\epsilon$-deformation in 
\beq
U_I^\epsilon (t) = T \exp \left( -i \int_{t_0(1-i\epsilon)}^t H_I(t^{\prime}) dt^{\prime} \right),
\eeq
as noted by \cite{Adshead:2008gk,Senatore:2009cf}. Nevertheless, Ref.  \cite{Kaya:2018jdo}, with further elucidation in our forthcoming work \cite{anotherBSpaper}, has shown that there is a perturbatively equivalent $\epsilon$ deformation, 
given by
\beq
U_I^\epsilon (t) = T \exp \left( -i \int_{t_0 }^t H^{\epsilon}_I(t^{\prime}) dt^{\prime} \right),
\eeq
where $H^{\epsilon}_I$ is the result of evaluating the interaction on an $\epsilon$-deformed field, 
\beq
\phi^{\epsilon} \equiv \phi \, e^{\epsilon t}.
\eeq 
With this  new prescription, 
\beq
\bra{\Omega} \mo \ket{\Omega} (t) = \bra{\rm BD'} U_I^{\epsilon\, \dag} (t) \, \mo \, U_I^\epsilon (t) \ket{\rm BD'}.
\label{eq:pertsetup}
\eeq
The advantage of this new formulation is that $U_I^\epsilon$ is unitary and our formal proof of Eq.~\ref{eq:pertex} now goes through.

By normal ordering the creation and annihilation operators and tracking nontrivial commutations in Eq.~\ref{eq:pertex}, we convert any perturbative contribution to the expectation value to an appropriate convolution of two-point functions.  The nontrivial task is to determine which two-point functions give the simplest formulation.  We know, for example, that with traditional time-ordered or ``in-out'' correlation functions, it is most straightforward to use two-point functions which are Feynman propagators.  It is well known that expectation values of the ``in-in'' type that interest us here have a more complicated structure and necessarily utilize more than one type of two-point function.  In fact, there are multiple equivalent approaches that use different sets of propagators, see for example Refs.~\cite{Jordan:1986ug,Musso:2006pt}.   Since our aim here is to identify the leading-log contributions, we will identify the basis of two-point functions that allows us to do so most efficiently.  Our starting point is the nested commutator expression for $\vev{\phi_{\rm Heis.}(t,x)^n}$, Eq.~\ref{eq:pertex}.  

\begin{figure}
\begin{center}
\includegraphics[width=12cm]{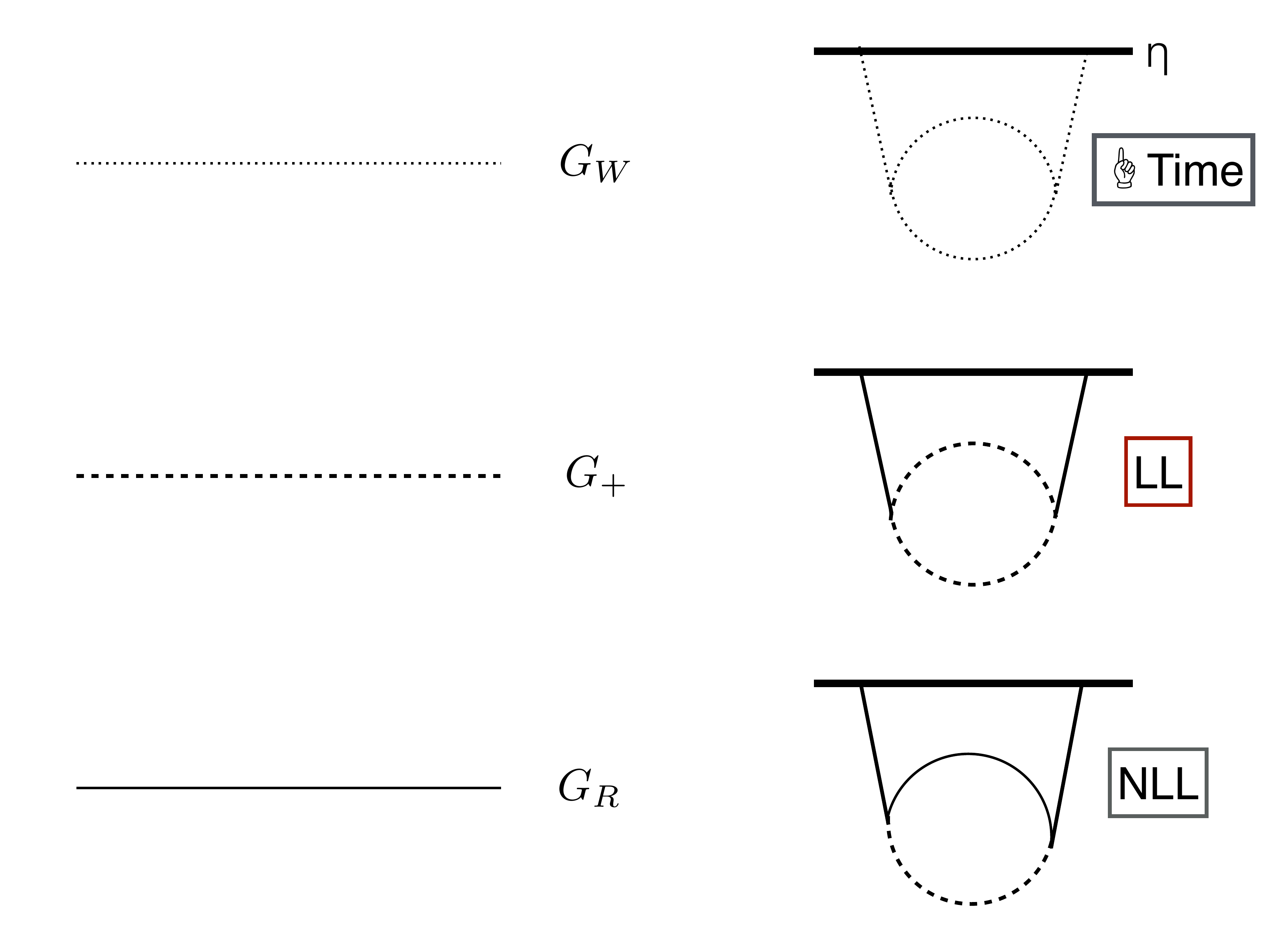}    
\end{center}
\caption{{\bf Left:} Our graphical notation for the various propagators of interest.  {\bf Right:} In a $\phi^3$ theory, we show second-order contributions to $\vev{\phi(\eta, \vec x) \phi(\eta, \vec y)}$, where the solid bar at the top of each diagram indicates the correlation time, $\eta$.  The top graph fixes a particular topology for this contribution, all of which will come from various contributions of Wightman functions and time orderings.  We can decompose the various contributing two-point functions purely in terms of $G_+$ and $G_R$.  The middle graph shows a nonvanishing contribution in this basis that contributes at leading-log level by having the minimum number of allowed $G_R$ factors and places them consistently with rules 1) and 2).  Lastly, the bottom graph is subleading-log as it contains more than the minimum number of $G_R$ terms. 
\vspace{0.025in}}
\label{fig:propagators}
\end{figure}
%

\subsection{Causality Constraints}

The "nested" form of the expectation value given in Eq.~\ref{eq:pertex} implies two causality-related
constraints  that any nonvanishing perturbative contribution must satisfy, as first noted in Ref. \cite{Musso:2006pt}. 
 There it is recognized that an in-in expectation value is most efficiently given by a combination of $G_R,\, G_+$, and simple Wightman functions, $G_W = \vev{\phi(x) \phi(y)}$, which arise from symmetrizing over more than two fields.    Given our focus in the current paper on the soft physics, we will see that we can exploit further simplifications to economically capture the leading infrared behavior in terms of just $G_R,\, G_+$. The two causality constraints are:
\begin{quote} \begin{itemize}
\item[1)] There is a Wick contraction across every comma in a commutator in Eq.~\ref{eq:pertex}.  Furthermore, Wick-contracting across a comma always yields at least one $G_R$. 
\end{itemize} \end{quote}
We give an example of this in Fig.~\ref{fig:propagators}.  Nonvanishing graphs have $G_R$ propagators touching every vertex and at least one touching the correlation point.  
If we broke the above rule 1), then after contracting, we would get a commutator of $c$-number functions, which would necessarily vanish.  
Schematically, we have
\be
\contraction{\Big \langle \Big [ \Big [ \ldots \Big [}{\phi}{(t,x)^n, }{H_I}
\contraction{\Big \langle \Big [ \Big [ \ldots \Big [ \phi_I(t,x)^n,  }{H_I}{(t_V) \Big ],}{\ldots}
\bcontraction{\Big \langle \Big [ \Big [ \ldots \Big [ \phi_I(t,x)^n,  H_I(t_V) \Big ], }{\ldots}{ , }{H_I}
\bcontraction[1.5ex]{\Big \langle \Big [ \Big [ \ldots \Big [ \phi_I(t,x)^n,  H_I(t_V) \Big ], \ldots , H_}{I}{(t_2) \Big ],}{H_I}
\Big \langle \Big [ \Big [ \ldots \Big [ \phi_I(t,x)^n,  H_I(t_V) \Big ], \ldots , H_I(t_2) \Big ], H_I(t_1) \Big ] \Big \rangle &\neq& 0 \nn \\[5pt]
\contraction{\Big \langle \Big [ \Big [ \ldots \Big [}{\phi}{(t,x)^n, }{H_I}
\contraction{\Big \langle \Big [ \Big [ \ldots \Big [ \phi_I(t,x)^n,  }{H_I}{(t_V) \Big ],}{\ldots}
\bcontraction[1.5ex]{\Big \langle \Big [ \Big [ \ldots \Big [ \phi_I(t,x)^n,  H_I(t_V) \Big ], \ldots , H_}{I}{(t_2) \Big ],}{H_I}
\Big \langle \Big [ \Big [ \ldots \Big [ \phi_I(t,x)^n,  H_I(t_V) \Big ], \ldots , H_I(t_2) \Big ], H_I(t_1) \Big ] \Big \rangle &=& 0,
\ee
where the absence of a Wick contraction across the comma before $H_I(t_2)$ in the second line causes the whole term to vanish.  Implicit in the first line is that every comma is contracted across.  We always get at least one $G_R$ from this because the commutator between operators $H_I$ or $\phi^n$ necessarily involves a minus sign.  Thus, we must get at least one $G_R$ to provide the sign.  Any contribution with fewer necessarily vanishes.  
\begin{quote} \begin{itemize}
\item[2)] A consequence of needing a contraction across every comma is that each vertex has a retarded propagator connecting it to either a vertex at a later time or the correlation point (the latest time, $\eta$, when the observable's expectation value is being taken).
\end{itemize} \end{quote}
This follows from the fact that the $H_I$ terms along with $\phi(t,x)^n$ appearing in Eq.~\ref{eq:pertex} are time-ordered.

We thus see that upon decomposing propagators into $G_R$ and $G_+$ the structure of diagrams is highly constrained by rules 1) and 2).   In particular, any Wightman function that appears in the full Wick contraction can be written as
\beq
G_W = \frac 1 2 ([ \phi(x), \phi(y) ] + G_+) \;\to\; \frac 1 2 (G_R + G_+),
\eeq
where we can trade the pure commutator two-point function for $G_R$ since the times in our operators are ordered by the integrals over time in the full correlation function ({\it cf.}~Eq.~\ref{eq:pertex}).  As we will see below, by soft power counting, we can drop any such $G_R$ factors that appear from $G_W$.  This means that only the $G_R$ that make up the causal skeleton (mandated by the causality constraints) are needed, and one can work entirely with the two propagators, $G_R$, $G_+$, for the leading description. 
 
\begin{figure}
\begin{center}
\includegraphics[width=8cm]{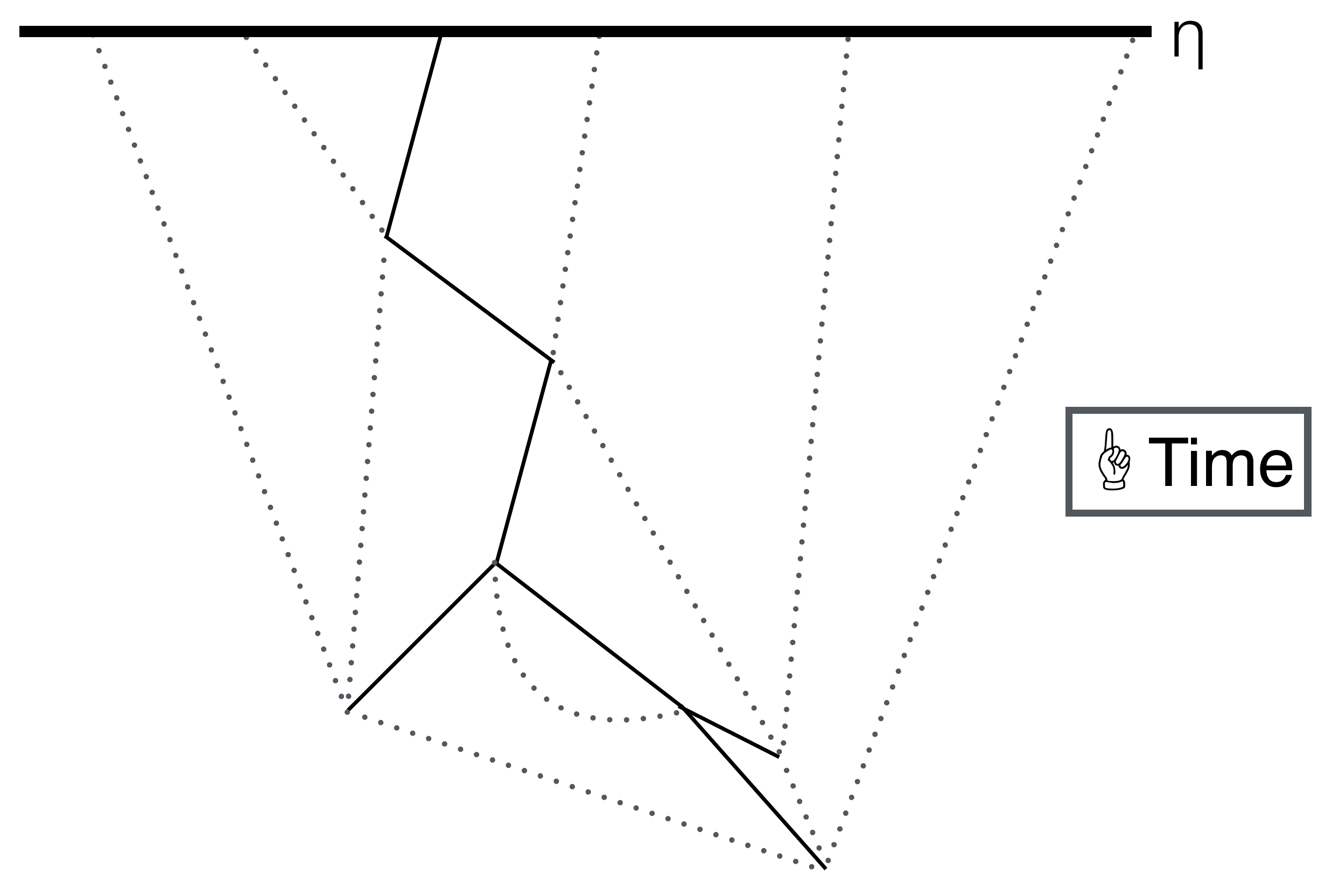}
\raisebox{0.025\height}{\includegraphics[width=8cm]{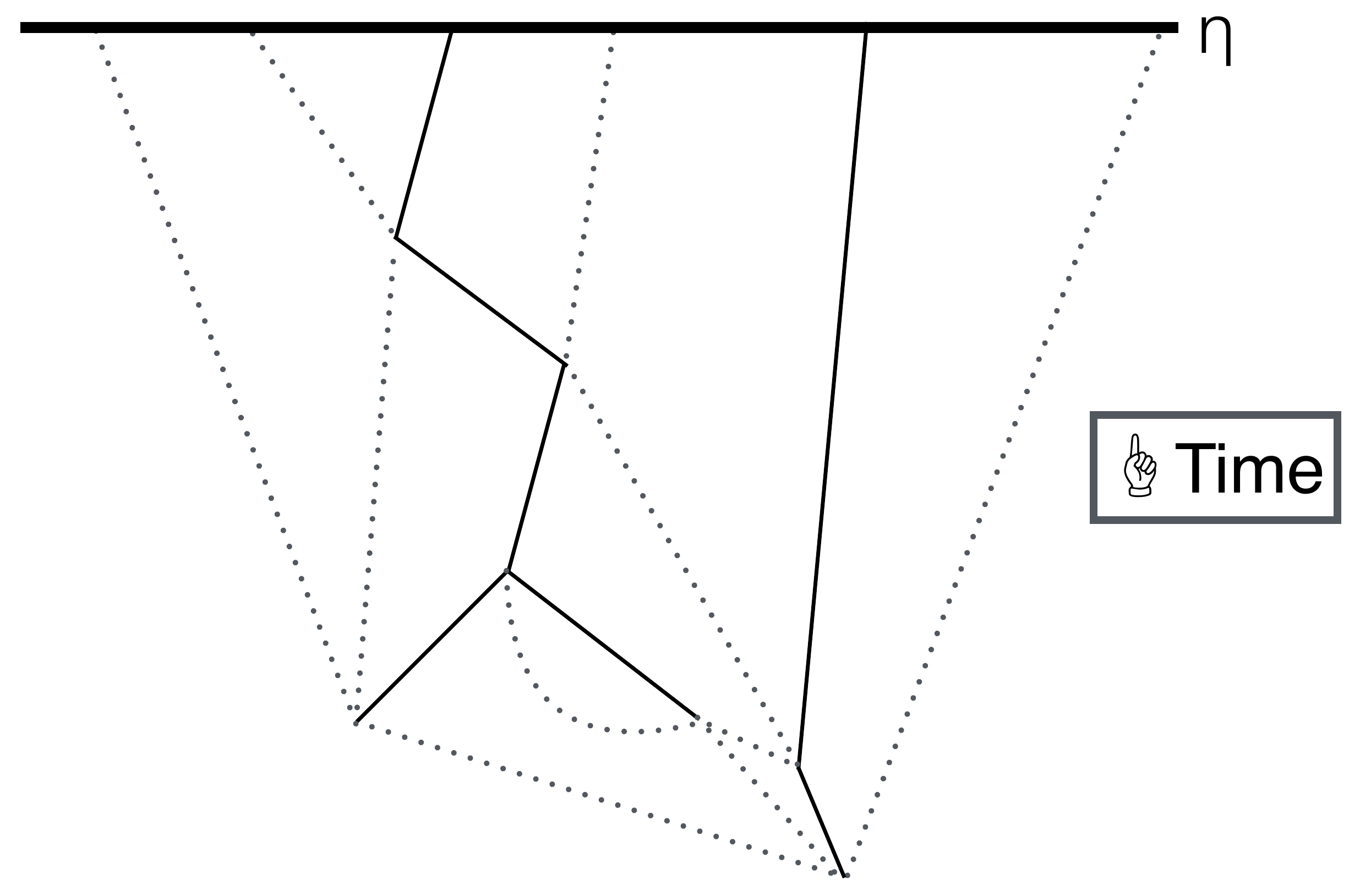}}
\caption{Examples diagrams of $\vev{\phi^6}$ evaluated at correlation time $\eta$ in $\lambda \phi^4$ theory at $\mo(\lambda^7)$.  For visual clarity we have separated out the six correlation points, though we generally take them to be spatially coincident, as well.  {\bf Left:} Illustration of the statement that if the subdiagram containing all $V$ retarded propagators is a connected tree that only touches the correlation point once, then it necessarily touches all $V$ vertices.  Solid lines are $G_R$ and dashed lines are $G_+$ propagators.  {\bf Right:} An example of the general situation with multiple disconnected subgraphs.  By the same argument, each one is a tree that touches one and only one external correlation point and contains a number of propagators equal to the number of vertices in the subgraph.}
\label{fig:onetreesub}
\end{center}
\end{figure}

We now wish to determine the properties of diagrams that have leading sensitivity to $\kir$.  Starting from Eq.~\ref{eq:momwightreg}, we can easily obtain the full momentum dependence for our propagators of interest.  Keeping the IR regulator implicit, we have
\be
G_R(k;\, \eta^\prime, \eta) &=&  \theta(\eta^\prime - \eta) \frac{-i \, H^2}{k^3} \Big[(1 + k^2 \eta \, \eta^\prime) \sin \left[ k(\eta^\prime - \eta) \right] - k(\eta^\prime - \eta) \cos \left[ k(\eta^\prime - \eta) \right] \Big],  \nn \\
G_+(k;\, \eta^\prime, \eta) &=& \frac{H^2}{k^3} \Big[ \cos \left[ k(\eta^\prime - \eta) \right] (1+ k^2 \, \eta^\prime \eta) + k \, (\eta^\prime - \eta) \sin \left[ k(\eta^\prime - \eta) \right]  \Big ].
\label{eq:pmprops}
\ee
The intuition for why there should be such a nontrivial structure for the in-in Feynman diagrams just comes from the soft $k$ scaling of the Green's functions,  
\be
G_R &\sim& k^0 \nn \\
G_+ &\sim& k^{-3}.
\label{eq:gscaling}
\ee
It thus appears that for leading sensitivity to the IR cutoff, we should economize on $G_R$ propagators, though we stress that some number of them are needed for causality.\footnote{In analyzing effects of the Higgs boson on inflation and the appropriate renormalization scale for the Higgs quartic, $\lambda(\mu)$,  \cite{Kearney:2015vba} noted that the IR sensitivity is enhanced in their calculations for contributions coming from $G_+$ propagators.}  Taking only the minimal number of $G_R$ propagators needed for a nonzero contribution to $\vev{\phi^n}$ forces an important topological constraint on the subdiagrams made entirely of $G_R$ lines.  If we need to touch all $V$ vertices and at least one external correlation point, with every vertex connected by a $G_R$ to a vertex at a later time or a correlation point, then the $G_R$ subdiagrams are all trees that touch one and only one external correlation point, as we illustrate in Fig.~\ref{fig:onetreesub}.\footnote{The idea that one can calculate in-in diagrams as tree subgraphs of $G_R$ propagators contracted by $G_+$ and $G_W$ goes back to Ref.~\cite{Musso:2006pt}, albeit in an analysis of the general structure of perturbation theory.  In a discussion of secular growth from IR divergences in single-field inflation, Ref.~\cite{Urakawa:2009my} also noted the tree subgraphs of $G_R$ terms, as well as the different soft scaling of $G_R$ from that of the $G_W$ propagators they used to contract the trees to make full diagrams.  However, their approach did not extend to all orders or include resummation, which form the focus of the present paper.}  The minimal number of $G_R$ propagators needed to "straddle" the vertices and correlation point in this way is precisely $V$.
%

\subsection{The Fast Track to Leading-Log}
\label{subsec:llexp}
  
Following the plausible intuition that we should economize on $G_R$ lines, in favor of the more IR-singluar $G_+$ lines, in order to maximize sensitivity to $\kir$, 
we have seen that an arbitrary (multi-loop) diagram has an important tree-level $G_R$ substructure,  "dressed" by $G_+$ lines to form loops. 
We will now determine the degree of IR sensitivity to $\kir$ of any such graph, in a simple graphical way. It is possible that this intuition might fail when loop momenta become highly virtual and hard so that there is no obvious preference for $G_+$ over $G_R$, although one might expect such hard contributions to be subleading in IR sensitivity. In Appendix B, it is proven rigorously that this indeed the case, so the leading IR sensitivity does require the minimal tree-substructure of $G_R$. In Section \ref{sec:fp}, we will detail how this ``loops from trees'' construction of De Sitter perturbation theory is a direct manifestation of the semiclassical nature of the underlying dynamics.
  
Starting with a completely general graph with $V$ vertices and $P$ propagators, contributions to coincident $n$-point functions take the following form,
\be
\vev{\phi(\eta,0)^n} \Big |_{\lambda^V} &\sim& \lambda^V \, \int_{1/\kir}^{\eta} \frac{d\eta^{(V)}}{(H \eta^{(V)})^4} \ldots
\int_{1/\kir}^{\eta^{(2)}} \frac{d\eta^{(1)}}{(H \eta^{(1)})^4} \, \int_{\kir}  \frac{d^3k_1}{(2\pi)^3} \ldots \int_{\kir} \frac{d^3k_{P \!-\! V}}{(2\pi)^3}  \nn \\
&&\times \prod_{m=1}^{P} G \left(k_m;\, \eta^{(m_1)},\eta^{(m_2)} \right).
\label{eq:phischem}
\ee
Recall that above Eq.~\ref{eq:momwightreg}, we have discussed the origin of the initial time and infrared momentum cutoffs.  We leave any labels off of $G$ to start.  We note that in every diagram, there will be $P-V$ undetermined momenta, which is most easily seen by thinking of the correlation point as a vertex and then doing the usual counting of loop momenta.  Thus, $k_m$ in the propagator lines will generally be some linear combination of loop momenta.  The $\kir$ dependence enters in two ways.  Firstly, it is the comoving cutoff on our loop integrals, and secondly it sets the initial times on our vertex integrals.  To determine $\kir$ scaling, we will approximate $G_R$ and $G_+$, starting by keeping only their leading terms in the soft limit ({\it cf.}~Eq.~\ref{eq:gscaling}).  In Appendix \ref{app:closel}, we show how the higher-order terms in the momentum expansion do not change the leading $\kir$ power counting.

Rules 1) and 2) require that a retarded propagator, $G_R$, must touch every vertex and at least one external correlation point.  The minimal number of $G_R$ lines we can take is to have $V$ of them, with all $G_R$ subgraphs forming trees. The resulting contribution is
\be
\vev{\phi(\eta,0)^n} \Big |_{\lambda^V} &\sim& \lambda^V \, \int_{1/\kir}^{\eta} \frac{d\eta^{(V)}}{(H \eta^{(V)})^4} \ldots
\int_{1/\kir}^{\eta^{(2)}} \frac{d\eta^{(1)}}{(H \eta^{(1)})^4} \nn \\
&&\times  \int_{\kir} \frac{d^3k_1}{(2\pi)^3} \ldots \int_{\kir} \frac{d^3k_{P \!-\! V}}{(2\pi)^3}  \nn \\
&&\times \prod_{i=1}^{V} G_{R\, {\rm soft}} \left(\eta^{(i_1)},\, \eta^{(i_2)} \right) \, \prod_{j=1}^{P-V} G_{+\, {\rm soft}} \left(k_j \right). 
\label{eq:phiso}
\ee

To proceed, it is useful to divide the $\eta$ integration in Eq.~\ref{eq:phischem} into one region where {\it all} the times are strongly ordered ($|\eta_{\rm earlier}| \gg |\eta_{\rm later}|$) and other regions where at least some of the times are comparable, $\eta_{\rm earlier} \sim \eta_{\rm later}$.
 In Appendix \ref{app:closel}, we will show that all the non-strongly-ordered contributions are parametrically subleading in sensitivity to $\kir$, and can therefore be dropped to get the leading estimate. (Indeed, we will get a good intuition for why this is the case just by studying the strongly-ordered regime.)
Under the assumption of strongly-ordered times, we can use the further approximation for $G_R$ in the soft limit,
\be
G_{R\, {\rm soft}}(\eta,\, \eta^\prime) &=& \theta(\eta^\prime - \eta) \frac{i\, H^2}{3}(\eta^3 - \eta^{\prime 3}) \nn \\
&\approx& \theta(\eta^\prime - \eta) \frac{i\, H^2}{3}\eta^3.
\label{eq:retapprox}
\ee
Similarly, we have
\beq
G_{+\, {\rm soft}}(k) = \frac{H^2}{k^3}.
\label{eq:plusapprox}
\eeq

As we see in Fig.~\ref{fig:onetreesub}, with only $V$ retarded propagators in the graph, each vertex time $\eta^{(i)}$ is the earliest time in one and only $G_R$ propagator.  
Given the strongly-ordered simplification of $G_{R\, {\rm soft}}$, this means that our integrand has one $\eta^{(i)\, 3}$ factor for each $i$.  These combine with the $\eta^{(i)\, -4}$ terms in the measure to give an overall $\eta^{(i)\, -1}$.  Thus, after plugging in the soft and strongly-ordered approximations for $G_R,\, G_+$, our contribution becomes  
\be
\vev{\phi(\eta,0)^n} \Big |_{\lambda^V} &\sim& \lambda^V \, \int_{1/\kir}^{A\, \eta} \frac{d\eta^{(V)}}{\eta^{(V)}} \ldots
\int_{1/\kir}^{A\, \eta^{(2)}} \frac{d\eta^{(1)}}{\eta^{(1)}}  \nn \\
&&\times \int_{\kir} \frac{d^3k_1}{(2\pi)^3 \, k_1^3} \ldots \int_{\kir} \frac{d^3k_{P \!-\! V}}{(2\pi)^3 \, k_{\pmv}^3} +\; ({\rm subleading}),
\label{eq:effgrexpl}
\ee
where $A$ is a modestly big number that enforces strong ordering of times.  There are $\pmv$ $G_+$ propagators, and there are $\pmv$ loop momenta.  Since all $G_R$ propagators form tree subdiagrams, every $G_+$ we add to these trees to build up the complete graph adds an undetermined loop momentum.  Thus, we can assign momenta such that only a single loop momentum flows through each $G_+$.  If we focus purely on the $\kir$ dependence of the integrated result, we see that each of the $V$ $\eta$ integrals and each of the $\pmv$ $k$ integrals contributes one power of $\log(\kir)$ for an overall $\log(\kir)^P$ scaling.  The fact that all the $1/k^3$ sensitivities coming from $G_+$ propagators have multiplicatively factorized and are in one-to-one correspondence with loop momenta precludes the possibility of overlapping IR divergences and a more divergent $\kir$ scaling.  By translation invariance (and the fact that $H$ can be scaled out of the problem), the only dimensionful parameter that can balance the $\log$ argument is the correlation time, $\eta$.  

One may worry about the upper limit of the $k$ integrals, as Eq.~\ref{eq:effgrexpl} naively looks UV divergent.  We note though, that from the full form of the $G_R,\, G_+$ propagators given in Eq.~\ref{eq:pmprops}, in the limit of strongly-ordered times, the rapidly oscillating trigonometric factors will cut off the $k$ integrals beyond $k_j \sim 1/\eta_{\rm earliest, j}$, where $\eta_{\rm earliest, j}$ is the time of the earliest vertex that $k_j$ flows through.  However, the exact mapping of the UV cutoffs, $1/\eta_{\rm earliest,\, j}$, into the $\eta^{(i)}$ integration variables does not matter since $\int dx \log(x)^n/x \sim \log(x)^{n+1}$.  Each $\eta$ integral therefore still contributes a single $\log$.
We thus get our leading-log result for an arbitrary graph with $V$ vertices and $P$ propagators,
\be
\vev{\phi(\eta,0)^n} \Big |_{\lambda^V} &\sim& \lambda^V \, \prod_{i=1}^{V} \int_{1/\kir}^{A\, \eta^{(i+1)}} d\eta^{(i)} \, \eta^{(i)\, -1} \, 
\prod_{j=1}^{\pmv} \log(\kir \, \eta_{\rm earliest,\, j}) ,  \nn \\
&\sim& \lambda^V \log(\kir \, \eta)^{P},
\label{eq:llscaling}
\ee
In Appendix \ref{app:closel}, we consider the deviations from the approximations in getting to Eq.~\ref{eq:llscaling}, along with the general case of having more than $V$ retarded propagators, and show that they always have subleading dependence on $\kir$.  Intuitively though, we can already see that violating strong ordering kills one of the large logs from a time integral, the maximal number of large logs arising from fully hierarchical distribution of times. 

This analysis shows us that perturbation theory is predicting its own demise, with each graph contributing at $\log(\kir \eta)^P \sim (t-t_0)^P$.  The presence of $t_0$ in this expression reveals a possibly surprising sensitivity to the details of the start of inflation, despite the intuition that inflation is an efficient eraser of the past.  The dependence on $t$ shows that the problem only grows worse with time.  To fix the problems of perturbation theory, we will need to go beyond it.  Fortunately, the nonperturbative insight to do so lies within perturbation theory itself.

\section{Semiclassicality and First-Orderness}
\label{sec:scfo}

We have shown above in Section \ref{sec:llcorr} that to all orders in perturbation theory the leading-log contributions to $\langle \phi(x)^n \rangle$ are given by diagrams where the retarded propagators form tree-shaped subdiagrams, each of which touches one and only one external correlation point.  Furthermore, these tree subdiagrams touch every interaction vertex at least once.  We are led to associate these tree subdiagrams of retarded propagators with some sort of classical perturbation theory.  This leaves two questions: 1) What is the classical theory in which we are doing perturbation theory? 2) How do we recover a fully quantum correlation function, $\langle \phi(x)^n \rangle$, with this classical input?

\begin{figure}[htbp]
\begin{center}
\includegraphics[width=12cm]{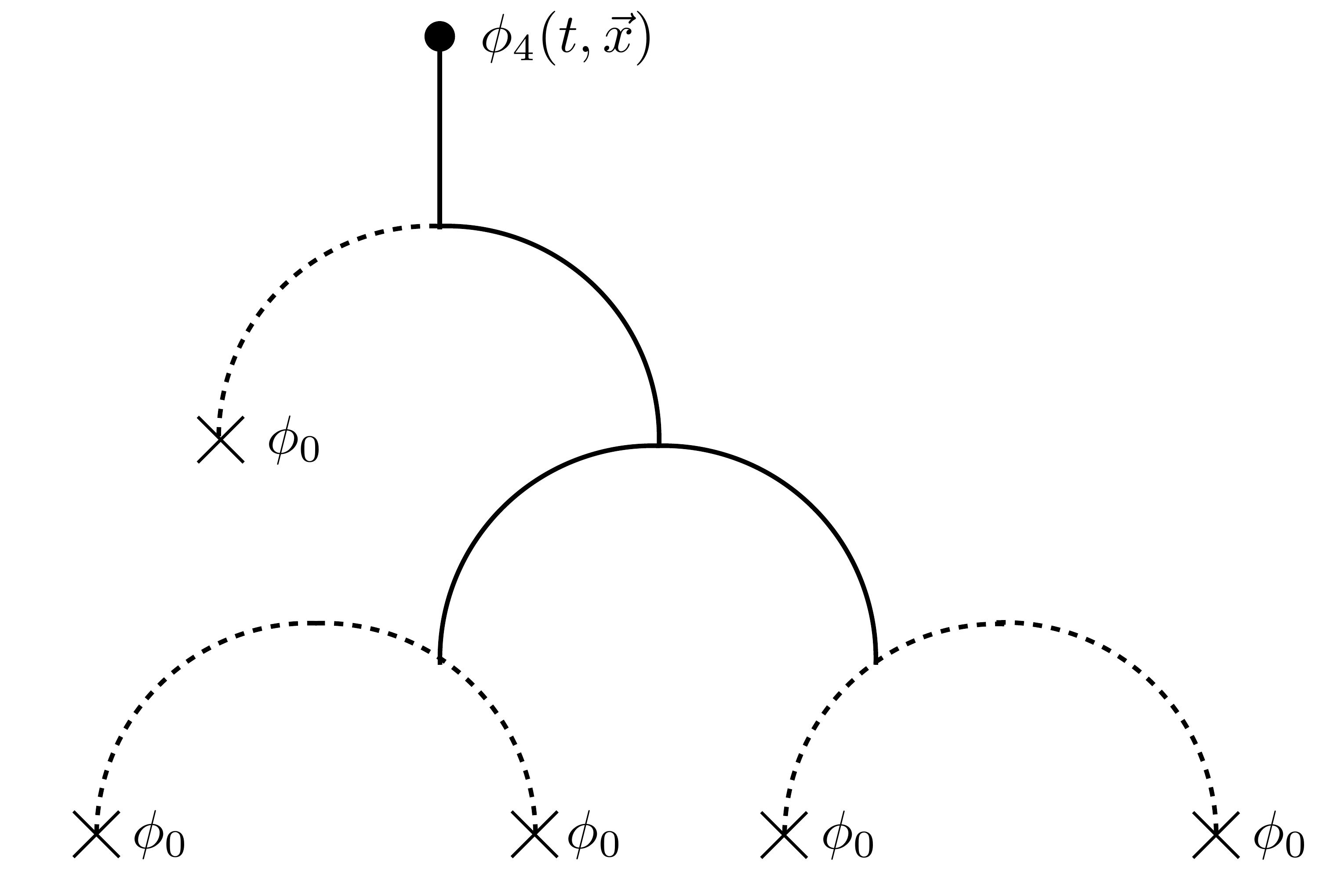}
\caption{Example of a classical perturbation theory Feynman diagram.  In this case, it is the fourth-order perturbative correction, $\phi_4(t, \vec x)$, to the full field solution, $\phi(t, \vec x)$, in classical $\phi^3$ theory.  Solid lines are retarded propagators, $G_R$, which get convolved with each other, and ultimately the free solutions $\phi_0(t_i, \vec x_i)$.  To make contact with our quantum field theory diagrams, the free field insertions are depicted as dashed lines leading to a ``$\times$''.  If one were doing classical field theory, one would need to use initial/boundary conditions or a specific inhomogeneous source function to give a particular $\phi_0$ for a corrected solution.}
\label{fig:classpert}
\end{center}
\end{figure}

The answer to question 1) is straightforward.  The retarded Green's functions, expanded in the limit of soft momenta, $G_{R\, {\rm soft}}$, are precisely those of the zeroth-order equation,
\beq
\partial_t \phi \equiv \dot \phi = 0,
\label{eq:zeom}
\eeq
in the following sense.
In Section \ref{sec:llcorr}, we showed that one ingredient for obtaining the leading-log correlation functions is the propagator,  $G_R$, expanded in both the soft and strongly-ordered limits $|\eta^\prime| \ll |\eta|$,
\beq
G_R(\eta,\eta^\prime;k) \approx \theta(\eta^\prime - \eta) \frac{i H^2}{3} (\eta^3 - \eta^{\prime\, 3}) \approx \theta(\eta^\prime - \eta) \frac{i H^2}{3} \eta^3.
\label{eq:gretapp}
\eeq
One can take the final expression in Eq.~\ref{eq:gretapp}, and use it as an input to reverse engineer the equation of motion for which it is the retarded Green's function.  We see that it solves the first-order, gradient-less approximation, Eq.~\ref{eq:zeom}, to the complete equation of motion (Eq.~\ref{eq:eom}), 
\be
3H \partial_{t^\prime} G_R(\eta^\prime,\eta;k) = \frac{1}{a^3} \delta(t^\prime - t) \nn \\
-3H^2 \eta^\prime \partial_{\eta^\prime} G_R(\eta^\prime,\eta;k) = (H\eta^{\prime})^4 \delta(\eta^\prime - \eta).
\label{eq:zerothapp}
\ee
Thus, by using the approximate Green's function for superhorizon modes ($k \eta \ll 1$), we are working in the first-order, gradient-less approximation to the classical theory.  Our leading-log result included the interactions from $V(\phi)$ to all orders.  Thus, we can now understand the tree, solid-line subdiagrams of $G_R$ propagators as perturbatively solving the first-order equation of motion,
\beq
\dot \phi = -\frac{1}{3H} V^\prime(\phi),
\label{eq:foeom}
\eeq
taking $G_R$ from the solution of the zeroth-order Eq.~\ref{eq:zerothapp}.  Fig.~\ref{fig:classpert} shows how the classical equation, $\dot{\phi} = -\frac{1}{3H} V^\prime(\phi)$ is solved perturbatively, giving a diagrammatic expansion in terms of $\phi_0$, the solution to the noninteracting Eq.~\ref{eq:zeom}.  

Our subdiagrams therefore {\it would} be a perturbative solution to the classical field theory with an equation of motion given by Eq.~\ref{eq:foeom} if we inserted 
zeroth order classical solutions, $\phi_0$, on the terminal branches.  We have however identified these subgraphs in the leading-log result of a complete QFT calculation, 
in which the standard appearance of such classical $\phi_0$ on terminal branches of retarded trees is instead replaced pairwise by the quantum propagator $G_+$. In this sense, the zeroth order $\phi_0$ ``seed'' for the nonlinear classical perturbation expansion is drawn from a quantum distribution with two-point correlation $G_+$.  
Thus, we see technically at the diagram level the qualitative physics of the original Starobinsky formulation of Stochastic Inflation:  quantum noise, which can be treated consistently in perturbation theory and even approximated in leading order as a free theory, when sufficiently redshifted evolves by interacting, first-order classical dynamics.  In this way we see in practice how the oft-stated claim that ``superhorizon modes in De Sitter are semiclassical'' emerges.     

The equation that gives the evolution of these tree subdiagrams, Eq.~\ref{eq:foeom}, approximates the full classical equation of motion (Eq.~\ref{eq:eom}), but we have seen that it is sufficient to recover the leading-log result.  As a consistency check, we show in Appendix \ref{app:accel} that reinstating the acceleration term, $\ddot \phi$, is equivalent to adding an effective interaction.  We can therefore include the effect of acceleration as a perturbation to the first-order equation that the leading-log graphs solve.  The replacement is $\ddot \phi \to \frac{1}{9H^2} V^\prime(\phi) V^{\prime \prime}(\phi)$, but this new effective vertex always contributes at subleading log order because it is higher order in the original coupling, $\lambda$, without any extra logarithms.

\section{Log Resummation as Fokker-Planck Evolution}
\label{sec:fp}
  
\begin{figure}
\begin{center}
\includegraphics[width=12cm]{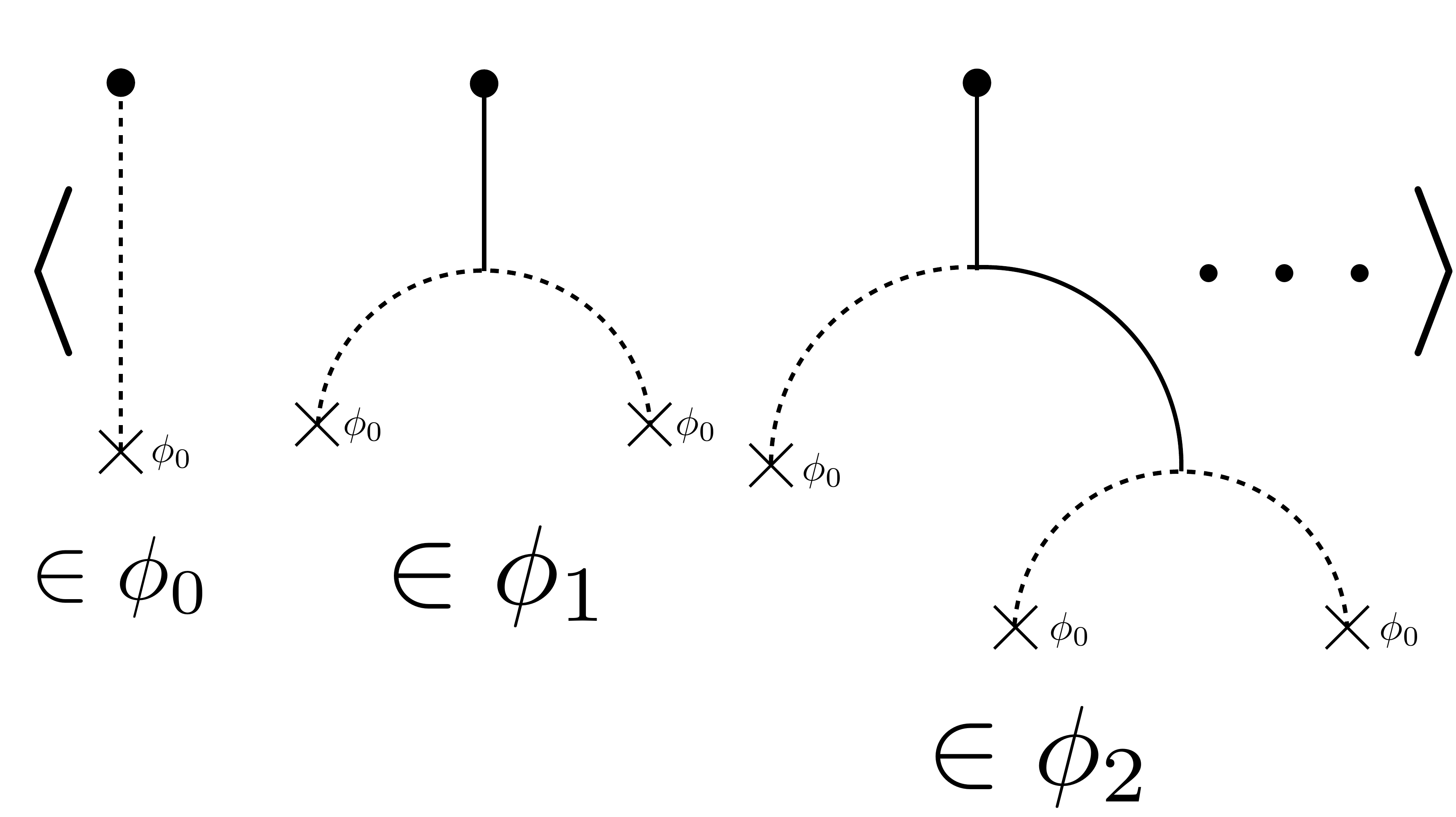}    
\end{center}
\caption{A schematic representation of Eq.~\ref{eq:qtosc}.  The graph labelled $\phi_0$ represents those contributions arising from the direct insertion of the free-field solution, $\phi_0$, which can be generalized to the inhomogeneous source term in Eq.~\ref{eq:inhomeom}.  The middle graph shows those evaluated at $\mo(\lambda)$ in classical perturbation theory, and the last graph is a representative of those at $\mo(\lambda^2)$.  There may be multiple contributions at each order (the $n_i$ in Eq.~\ref{eq:qtosc}) and at higher orders.  The interpretation of $\vev{}$ is that it joins up the dangling $\times$ factors of each classical solution pairwise in all possible combinations, evaluating them as the anticommutator propagator, $G_+$.}
\label{fig:vevd}
\end{figure}

Let us rewrite our diagrammatic expansion in the following, useful way.  For the contribution with $V$ vertices, we explicitly perform any Wick contraction that results in a retarded propagator, $G_R$, but we momentarily leave undone those that give anticommutator propagators, $G_+$:\footnote{Under the full in-in decomposition found in \cite{Musso:2006pt}, there are Wightman functions in addition to $G_+,\, G_R$.  However, in the soft limit, we have $G_W \to \frac 1 2 G_+$, allowing us to simplify our basis.  Furthermore, tracking these factors of $\frac 1 2$ is not needed in detail at the order in which we are working.} 
\be
\vev{\phi(t,x)^n}\Big |_{\lambda^V} &=& \sum_{\substack{{\rm perms.}\\ \sum_i n_i = n}}  \vev{\phi_0(t,x)^{n_0}\, \phi_1(t,x)^{n_1}\, \phi_m(t,x)^{n_m}}, \; {\rm where} \nn \\
\phi_1(t,x) &=& -\int \frac{d^4y}{(H \eta)^3} G_R(x,y) V^\prime(\phi_0(y)) \nn \\
\phi_2(t,x) &=& -\int \frac{d^4y}{(H \eta)^3} G_R(x,y) V^\prime(\phi_0(y)+ \phi_1(y)) \big |_{\mo(\lambda^2)}, \;\; {\rm etc.,}
\label{eq:qtosc}
\ee
and $1+2+\ldots m = V$.  The $\mo(\lambda^2)$ in the expression for $\phi_2$ just means that we expand out the argument to second order. 
In general, $\phi_j$ is the $j$th order piece of the classical solution, given zeroth order solution $\phi_0$.
 Thus, since $V^\prime$ already contains an explicit coupling, at $\mo(\lambda^2)$ this equation will only contain the contribution from $V^\prime$ with a single $\phi_1$ and all the rest $\phi_0$.  The graphical interpretation of Eq.~\ref{eq:qtosc} ({\it cf.}~Fig.~\ref{fig:vevd}) takes our original Feynman diagram, but cuts every $G_+$ line and writes $\phi_0(x_V)$ on each newly exposed end, where $x_V$ is the spacetime location of the vertex to which it attaches.  Cutting all the $G_+$ lines in this way just leaves us with a set of classical perturbation theory diagrams for each external $\phi(t,x)$ in the correlator, except those where the field at the correlation point is itself a $\phi_0$ factor.  We can get back to the full correlation function by performing the remaining $G_+$ Wick contractions of the $\phi_0$s in Eq.~\ref{eq:qtosc}.
Since these are symmetric, the ordering of our fields in the expectation value does not matter.  In this way, we can cover a topology with arbitrary loops from a fundamental basis of classical, tree graphs.  Eq.~\ref{eq:qtosc} gives precisely the leading-log contribution to $\vev{\phi(t,x)^n} \big |_{\lambda^V}$ since the explicit sum contains all possible ways of getting trees of retarded propagators that satisfy our causality constraints ({\it cf.~Theorem} at the start of Section \ref{sec:llcorr}).  Performing the anticommutator Wick contractions then provides all possible ways of linking these trees to each other.  

We can thus interpret the expectation value at fixed order in perturbation theory in terms of a classically-evolved, interacting field theory convolved with quantum, symmetric, two-point expectation values.  Our concern is what happens to the $n$-point function at late times.  However, it is intractable to quantify this behavior with direct calculation beyond $t - t_0 \gtrsim \lambda^{-2/m}$, for a potential $V(\phi) \propto \lambda \, \phi^m$, in light of the breakdown of perturbation theory.  

It is possible though, to write down a simple update equation for the $n$-point function.   Each of the expanded $\phi$ terms in Eq.~\ref{eq:qtosc} is just perturbatively solving the classical equation of motion, which we determined in Section \ref{sec:scfo} is approximately first-order and gradientless, $\dot{\phi} = -\frac{1}{3H} V^\prime(\phi)$.  We further know that quantum effects shift the zeroth-order equation, $\dot \phi = 0$, into an inhomogeneous equation, $\dot \phi = \dot \phi_0$, where the RHS is only known from a distribution.  The quantum expectation values $G_+$ gives us a nontrivial $\dot \phi_0$ because the $\phi_0$ terms have an anticommutator two-point function with nontrivial time dependence, $\partial_t \vev{\{\phi_0(x),\phi_0(y)\}} \neq 0$. We will see below that this time dependence has to be treated carefully. We can modify the classical equation of motion so that it does not constrain $\phi_0$ at all, in preparation for a proper quantum treatment of $\phi_0$, 
\beq
\dot{\phi} = -\frac{1}{3H} V^\prime(\phi) + \dot{\phi}_0.
\label{eq:inhomeom}
\eeq
Clearly, at zeroth order this reads trivially $\dot{\phi}_0 = \dot{\phi}_0$. Since diagrammatically the $\phi_0$ are pairwise contracted into $G_+$ (Eq.~\ref{eq:pmprops}), which is also independent of the coupling, $\dot{\phi}_0$ on the right-hand side is truly zeroth order, and therefore this modification does not enter the higher orders of the perturbative iteration. Thus the sole effect of this modification is for the classical equation to say nothing about $\phi_0$ itself, but to allow us to build up the nonlinear classical solution  $\phi(\phi_0)$ in terms of a given general $\phi_0$.

Together, Eqs.~\ref{eq:qtosc} and \ref{eq:inhomeom} allow us to write
\beq
\frac{d}{dt} \vev{\phi(t,x)^n} = \vev{\frac{d}{dt} \left( \phi(t,x)^n \right)} = -\frac{n}{3H} \vev{\phi(t,x)^{n-1} \, V^\prime[\phi(t,x)]} + n \vev{\phi(t,x)^{n-1} \dot{\phi_0}},
\label{eq:updatephizero}
\eeq
where in the first equality we have used the commutation of the time derivative with taking the quantum expectation value.  We can see this in the interaction picture as our bra and ket vector are evolved by $\exp[\mp i \int^t H_I (t^\prime) dt^\prime]$.  Acting on these exponentials with the time derivative just gives a commutator $\left[ H_I(t),\, \phi(t,x) \right]$, which vanishes if $H_I$ is purely a function of $\phi$, as both terms are at equal times.
The first term in Eq.~\ref{eq:updatephizero} provides the ``drift,'' and determines how the correlation functions change in the presence of the force derived from the potential $V[\phi(t,x)]$.  These interactions appear in the classical evolution of the field.  The second term is intrinsically quantum mechanical, as quantum fluctuations provide the nonzero value of $\dot{\phi}_0$ via $G_+$ contractions.  It is commonly known as the ``diffusion'' term since it describes the random fluctuation of scalar modes as they redshift into the regime where interactions become important.

Naively, the diffusion term is subdominant in the soft limit we are considering, because the soft limit of $G_+$ is time-independent, so that $\dot{\phi}_0 = 0$ within a soft $G_+$ contraction. Indeed, this is the case if  $\phi(\phi_0)$ is $G_+$-contracted with a $\phi_0$ in a typical interaction vertex at some earlier time, as further discussed below Eq.~\ref{eq:etacrxn}.  However, there is a subtle type of contribution we get from the $\dot{\phi}_0$ term contracting with another free $\phi_0$ in contributions where the latter also appears 
at the {\it correlation time} (as one of the $\phi_0$ appearing in the first line of Eq.~\ref{eq:qtosc}).  This does survive at leading-log. Because we are studying coincident point correlators, both the $\dot{\phi}_0$ and $\phi_0$ are at this coincident correlator point and it is inconsistent to take the conjugate momentum to be soft. Instead the coincident limit of the real-space $G_+$ suffers an ultraviolet divergence. Fortunately, as long as this divergence is regularized at a physical scale (and ultimately renormalized), the time dependence needed to compute $\vev{\phi_0(\eta, x) \dot{\phi}_0(\eta, x)}$ follows simply from the scale-factor conversion between the physical and comoving cutoff. 
As an explicit example, one can regularize the spatially coincident $G_+(\eta,\, \vec 0)$ by imposing a physical UV cutoff in momentum space,\footnote{In Appendix \ref{app:closel} we discuss why a hard cutoff on spatial momentum is legitimate for coincident correlation functions.} 
\beq
G_+(\eta,\, \vec 0) = \int^{\Lambda/H\eta}_{\kir} \frac{d^3 k}{8\pi^3} \frac{H^2(1+k^2 \, \eta^2)}{k^3},
\label{eq:coincg}
\eeq
which then leads to the cutoff-independent (renormalization point independent) result
\beq
\vev{\phi_0(\eta, x) \dot{\phi}_0(\eta, x)} = \frac{1}{4}\dot{G}_+(\eta,\, 0) = \frac{H^3}{8\pi^2}.
\label{eq:dotg}
\eeq
Furthermore, losing a factor of $\log(\kir \, \eta)$ on the RHS of the update equation, \ref{eq:updatephizero}, matches the explicit loss on the LHS.  

We can now simplify our $\dot{\vev{\phi^n}}$ update equation to give all leading-power contributions in terms of simple $\phi^n$ expectation values,
\beq
\frac{d}{dt} \vev{\phi(t,x)^n} = -\frac{n}{3H} \vev{\phi(t,x)^{n-1} \, V^\prime[\phi(t,x)]} + \frac{n(n-1) \, H^3}{8\pi^2} \vev{\phi(t,x)^{n-2}}.
\label{eq:updatephi}
\eeq
The $n(n-1)$ factor on the diffusion term has arisen from the combinatorics of pairing up two fields of the $n$ and replacing them with $\vev{\phi_0 \dot{\phi}_0}$.  We have already discussed the physical interpretation of the drift term as accounting for the evolution in the presence of the potential.  As we have seen, this dynamics is given by classical perturbation theory convolved with the quantum distribution for $\phi_0$. With the second term on the RHS, we include the change in $\vev{\phi^n}$ due to the presence of a noninteracting, quantum two-point fluctuation that occurs in the intervening time-step.  Since we can obtain the leading-log contribution to any correlation function by expanding in the soft limit, we can then interpret the $\vev{\phi^{n-2}}$ term in Eq.~\ref{eq:updatephi} as accounting for the redshifting of free hard modes into the soft region.  This is the sense in which it is giving diffusion. 

We see in Eq.~\ref{eq:updatephi} that the leading-log time evolution of coincident expectation values is ultralocal in space (by causality on superhorizon length scales).  The value of $\vev{\phi(t,x)^n}$ at later times only depends on higher and lower-point correlation functions at the same spacetime point, and has no spatial derivatives.  Furthermore, by spatial translation invariance, the expectation values cannot depend on $x$, and can thus be written, $\vev{\phi^n}(t)$.  We can thus replace the full quantum field, $\phi(t,x)$ with a 1D variable, $\phi$, whose expectation values can be computed by integrating against a time-dependent generating function, $p(\phi,t)$,\footnote{In Section \ref{sec:conc}, we sketch how $p(\phi,t)$ can be obtained from the QFT wavefunctional.}
\beq
\vev{\phi^n}(t) = \int d\phi \, p(\phi, t) \, \phi^n.
\label{eq:phidef}
\eeq
Now we show that $p(\phi, t)$ satisfies a Fokker-Planck equation,
\beq
\dot{p}(\phi, t) = \frac{1}{3H} \partial_\phi [V^\prime(\phi) \, p(\phi, t)] + \frac{H^3}{8\pi^2} \, \partial^2_\phi \, p(\phi, t).
\label{eq:theheroenters}
\eeq
One consistency check we note immediately is that in the absence of interactions, we just have a diffusion equation.  The resulting two-point function grows linearly with physical time (logarithmically with $\eta$), as we found for the coincident propagator, $\vev{\phi^2} \sim t$.
By integrating this against $\phi^n$, we see that we reproduce Eq.~\ref{eq:updatephi},
\be
\frac{d}{dt} \vev{\phi^n} \,=\, \int d\phi \, \dot{p}(\phi, t) \, \phi^n &=& \int d\phi \,  \frac{1}{3H} \left( \partial_\phi [V^\prime(\phi) \, p(\phi, t)] \right)\phi^n \,+\, \int d\phi \, \frac{H^3}{8\pi^2} \, \left( \partial^2_\phi \, p(\phi, t) \right) \phi^n \nn\\
&=& \int d\phi \,  p(\phi, t) \, \frac{-n}{3H}  V^\prime(\phi) \, \phi^{n-1} \,+\, \int d\phi \, p(\phi, t) \, \frac{ n(n-1) \, H^3}{8\pi^2} \phi^n \nn \\
&=& -\frac{n}{3H} \vev{\phi^{n-1} \, V^\prime(\phi)} + \frac{n(n-1) \, H^3}{8\pi^2} \vev{\phi^{n-2}}.
\label{eq:updatefromfp}
\ee

The power of Eq.~\ref{eq:theheroenters} comes from the fact that the generating function, $p(\phi,t)$, must reproduce the flow equation we derived to all orders in perturbation theory, but its solutions are not confined to the perturbative regime.  In particular, for a general $\lambda  \phi^m$ theory with $m$ even, using $p(\phi,t)$ parametrically extends the regime of computability for $\vev{\phi^n}$ up to times when the leading-log perturbations are large ($t \sim \lambda^{-2/m}$).  In this sense, we have resummed the leading logs to be trustworthy and dominant in a regime outside fixed-order perturbation theory. This is our rigorous central technical result.

It is tempting to connect this all-orders summation to the familiar case of renormalization group, but any detailed correspondence remains obscure at present.  One may also conjecture that NLL corrections may also be resummed, and that they will continue to be subleading to the LL effects even after resummation.  While a derivation of this fact awaits future work, one can nonetheless examine the Fokker-Planck solutions at arbitrarily late times {\it as if} such a statement were true.  As we show below, no obvious breakdown occurs in the formalism; we can thus gain insight into what plausible values of very-late time DS expectation values will be.      

We note that the Fokker-Planck equation (Eq.~\ref{eq:theheroenters}) can be written as a continuity equation,
\beq
\dot{p}(\phi, t) + \partial_\phi \, J(\phi, t) = 0,
\eeq
with probability current 
\beq
J(\phi, t) = -\frac{1}{3H} [V^\prime(\phi) \, p(\phi, t)] - \frac{H^3}{8\pi^2} \, \partial_\phi \, p(\phi, t).
\eeq
Thus, the quantity $\int d\phi \, p(\phi, t)$ is conserved, which allows us to interpret it as a probability. 

Following \cite{Starobinsky:1994bd}, we can recast Fokker-Planck in terms of a Euclidean Schr\"{o}dinger equation.  In particular, if we set $t_0=0$, solutions have the general form
\beq
p(\phi,t) \,=\, \exp \left[ -\frac{4\pi^2}{3H^4} V(\phi) \right] \, \sum_{n=0}^\infty \, a_n \Phi_n (\phi) e^{-\Gamma_n t},
\label{eq:psol}
\eeq
where the $\Phi_n (\phi)$ are the eigenfunctions of
\beq
\left[ -\frac{1}{2}\partial_\phi^2 + W(\phi) \right] \Phi_n(\phi) = \frac{4\pi^2 \, \Gamma_n}{H^3} \Phi_n(\phi).
\label{eq:eucschro}
\eeq
The Euclidean Schr\"{o}dinger potential is given by
\beq
W(\phi) = \frac{2\pi^2}{3H^4} \left[ \frac{4\pi^2}{3H^4} V^\prime(\phi)^2 - V^{\prime \prime}(\phi) \right].
\eeq
Thus, a potential $V(\phi)$ whose largest polynomial term is $\propto \phi^n$ and has a positive coefficient for $n\geq2$ is sufficient to guarantee a discretized spectrum.  Furthermore, we can rewrite Eq.~\ref{eq:eucschro} in the following way,
\beq
\frac{1}{2}\left[ -\partial_\phi + \frac{4\pi^2}{3H^4} V^\prime(\phi) \right] \left[ -\partial_\phi + \frac{4\pi^2}{3H^4} V^\prime(\phi) \right]^\dag\Phi_n(\phi) = \frac{4\pi^2 \, \Gamma_n}{H^3} \Phi_n(\phi),
\label{eq:eucschroalt}
\eeq
which makes manifest that the eigenvalues $\Gamma_n$ are nonnegative.  If the squared ground state eigenfunction, $|\Phi_0|^2$ is normalizable, then we have a zero eigenvalue with eigenfunction $\Phi_0(\phi) \propto \exp[-(4\pi^2/3H^4) \, V(\phi)]$.  We can therefore rewrite our general solution , Eq.~\ref{eq:psol}, as
\beq
p(\phi,t) \,=\, N \exp \left[ -\frac{8\pi^2}{3H^4} V(\phi) \right] + \exp \left[ -\frac{4\pi^2}{3H^4} V(\phi) \right] \, \sum_{n=1}^\infty \, a_n \Phi_n (\phi) e^{-\Gamma_n t},
\label{eq:psolzm}
\eeq
where $\Gamma_n >0$, and $N$ is a normalization factor.  The positivity of the $\Gamma_n$ means that solution flows to late-time fixed point.  Thus, $\lim_{t \tor \infty} \dot{p}(\phi,t) = 0$.  It is straightforward to see that plugging in $p(\phi,t) = N \exp[-(8\pi^2/3H^4) \, V(\phi)]$ solves the Fokker-Planck equation (Eq.~\ref{eq:theheroenters}) for $\dot{p}(\phi,t) = 0$.  Furthermore, we see that for nonpathological potentials, we are insensitive to the details of the initial condition on $p(\phi,t)$.  Thus, an initial state that differs perturbatively from the exact, free-theory Bunch-Davies, will flow to the same final distribution.

As an example, we can look at specific results for $V(\phi) = \lambda \phi^4/4!$.  Despite the badly-behaved perturbation series, correlation functions asymptote to finite values, {\it e.g.}~
\be
\lim_{t \rightarrow \infty} \vev{\phi^{2n}} &=& \frac{3^n \, H^{2n} \, \Gamma \left[ \frac 1 4 + \frac n 2 \right]}{\lambda^{n/2} \, \pi^n \, \Gamma \left[ \frac 1 4 \right]}, \nn \\
\lim_{t \rightarrow \infty} \vev{\phi^{2}} &=& \frac{3 \, H^2 \, \Gamma \left[ \frac 3 4 \right]}{\lambda^{1/2} \, \pi \, \Gamma \left[ \frac 1 4 \right]}.
\ee

The fact that our expectation values go like inverse powers of the coupling shows that dynamics at late times are controlled by fundamentally nonperturbative effects.  However, the breakdown in perturbation did not portend a deeper pathology in the theory.  Any remnant of our earlier departure from De Sitter geometry, parametrized by our IR cutoff, $\kir$, has disappeared.  Furthermore, at late times, we get the expectation value $\vev{V(\phi)} \sim H^4$, meaning that the interactions, which would classically drive the field to the origin, and the quantum fluctuations, which grow linearly in time in the absence of interactions, $\vev{\phi^2} \sim t$, reach an equilibrium wherein the field acquires a potential energy density, $H^4$, in accord with naive expectations from dimensional analysis for the expanding De Sitter spacetime.

\section{Conclusion \& Discussion}
\label{sec:conc}

Despite the apparent bad behavior of the perturbation series for $\vev{\phi^n}$, we see that the leading-log series in powers of $\log(\kir \eta)$ is resummed by the solution to the Fokker-Planck equation, and one obtains a healthy, bounded, De Sitter-invariant probability distribution $p(\phi,t) \sim \exp[-\# V(\phi)]$ at late times for any well-behaved potential.  It is natural to ask though, whether there is trouble lurking at NLL.  There are claims that even for a fixed De Sitter geometry, scalars are pathological \cite{Polyakov:2012uc}.  Given that we see no evidence of this sickness at the current, LL level, it is important to establish the health of the theory at higher orders.  The LL analysis of $\lambda \phi^4$ shows that at late times $V(\phi) \sim H^4$, which is the expected result as the energy density available to the expanding spacetime.  
What we see then at LL is that the large log growth is due to the diffusion effect, while the potential serves to ``contain" the field from diffusing without limit at late times, which is the key to resumming to a well-defined DS limit. It is therefore intuitively very plausible that this physical picture continues to hold at NLL, allowing it to be resummable and subdominant to the resummed LL, but we have not yet definitively established this. If this were {\it not} the case, then the subleading NLL graphs with $P$ propagators scaleing as $t^{P-1}$ 
would have the potential to overwhelm the perturbation series for sufficiently late correlator time.    

Thus, at the present stage, the result of our rigorous analysis extends the regime of calculability until times sufficiently late that the NLL contributions can potentially become large.  For a $\lambda \phi^4$ theory, the LL+NLL contributions take the schematic form,
\beq
\vev{\phi^n} =  \sum_{n = 0} a_n (\lambda \, t^2)^n + b_n \, \lambda \, t (\lambda \, t^2)^{n}.
\eeq
Our analysis is then guaranteed to be trustworthy in the limit $t \rightarrow \infty, \lambda \rightarrow 0$, with $\lambda t^2$ fixed, so that all the LL terms survive (and are resummed by Fokker-Planck evolution) but NLL $\rightarrow 0$. If however subleading log contributions do indeed resum to remain subleading, then Fokker-Planck evolution gives the leading nonperturbative behavior for correlators for large $t$ and finite small $\lambda$.

To proceed to NLL and beyond, one can systematically improve our treatment of LL, which used a series of well-defined approximations.   The central results of Sections \ref{sec:llcorr} and \ref{sec:scfo} show how first-order, semiclassical analysis emerges diagrammatically from looking at the leading sensitivity in correlation functions $\vev{\phi^n}$ to the comoving infrared cutoff, $\kir$, or the final correlation time.  We also saw that the leading evolution is ultralocal, with gradient contributions dropping systematically.  It is therefore straightforward to generalize to correlation functions of fields at an arbitrary number of spatial points, but all still coincident in time.  To compute at noncoincident times will be less trivial, but represents no qualitative challenge to the framework.  Indeed, the recent analysis of \cite{Gorbenko:2019rza}, which treats stochastic inflation as a similar systematic approximation within QFT, discusses these results as well.  

We show in Appendix \ref{app:accel} how one can include the field acceleration term in the equation of motion as a perturbation.  It gives an effective interaction vertex whose leading contributions are NLL in $\log(\kir \eta)$ power counting.  One could include the contributions from the gradient terms  by using the retarded Green's function for the first-order equation of motion that includes the $k^2/a^2$ term.  This will bring in subleading terms in the soft-momentum expansion of $G_R$ (Eq.~\ref{eq:pmprops}).  Higher-order terms in the momentum expansion of propagators $G_R,\, G_+$ also allow one to include the effects of perturbative interactions in the UV.  

Physically, the most nontrivial change at NLL comes from the inclusion of (long-distance) quantum entanglement effects.  In particular, the power counting analysis of Appendix \ref{app:closel} shows that one generically gets diagrams that either feature loops of $G_R$ propagators,\footnote{For in-in correlation functions, it is possible to arrange the time orderings such that a loop of retarded propagators does not automatically vanish.} or a line of $G_R$ propagators that connect one external correlation point to another.  Both of these effects spoil the semiclassical description of stochastic noise (short-distance quantum mechanical in origin) convolved with classical first-order evolution that ultimately led to the Fokker-Planck equation (Eq.~\ref{eq:theheroenters}).  Properly including the NLL effects will therefore likely necessitate a change in formalism. It could represent a point of contact between our approach and the recent, more manifestly path-integral framework of \cite{Gorbenko:2019rza}.  

In essence, our diagrammatic analysis rigorously justifies a heuristic derivation of Fokker-Planck dynamics from the wavefunctional/path-integral approach, which we sketch here. 
We see from the momentum space Wightman function that the poor $\sim 1/k^3$ IR behavior that underlies large logarithms in perturbation theory kicks in when the {\it physical} momentum falls well below $H$, $k \eta \ll 1$. Thus the log-enhanced interactions arise in this regime, while interactions for physical momenta $> H$ are not enhanced. We can therefore work in the leading approximation of neglecting interactions altogether for UV modes, $k \eta > 1$, and only retaining the IR log-enhanced interactions. This allows us to factorize the state wavefunctional, 
\begin{equation}
\Psi[\{\phi_k \}, t) \approx \Psi_{\rm UV, free~BD'} [\{ \phi_{k > 1/\eta} \}, t) ~ \times~ \Psi_{\rm IR, interacting}[\{ \phi_{ k<  1/\eta} \}, t). 
\end{equation}
We consider the correlators of interest in position space to be suitably coarse-grained over Hubble patches so that they are expressible in terms of the soft momentum modes alone, 
$\phi_{k < 1/\eta}$, and therefore only require knowledge of $\Psi_{\rm IR}$. The probability distribution functional for the coarse-grained $\phi$ field is then $|\Psi_{\rm IR}|^2$, 
 which expressed in position space we will denote by $P[\{\phi(\vec x) \}, t)$. 
This in turn can be reduced down to the probability function $p(\phi, t)$ at a single spatial point introduced in Eq.~\ref{eq:phidef}, say at the origin, $\phi \equiv \phi(\vec{x} = \vec{0})$, by integrating $P$ over all possible $\phi(\vec{x} \neq \vec{0})$.

First we consider a  completely free scalar theory, $\lambda = 0$, in which case the entire wavefunctional, both UV and IR, is free Bunch-Davies (BD$'$) in form. The free wavefunctional  $\Psi_{\rm IR}$ must be  exactly Gaussian in form. It follows that its square $P$, and further integral over $\phi(\vec{x} \neq \vec{0})$, $p$,   is also Gaussian. In order for Eq.~\ref{eq:phidef} to reproduce Eq.~\ref{eq:dotg}, we can deduce the specific late-time form, 
\begin{equation}
p \propto  e^{-2 \pi^2 \phi^2/(H^3 t)}.
\end{equation}
We can clearly see the diffusion of the field as a function of time here, and indeed the normalized $p$ satisfies the diffusion equation limit of  the Fokker-Planck equation, Eq.~\ref{eq:theheroenters}, in the absence of any potential interactions, $V=0$. The root of this effect is simply the redshifting of comoving momentum modes through the physical cutoff scale, as seen in Eq.~\ref{eq:coincg}. In the above UV-IR factorization, this effect is simply due to the fact that comoving modes in the UV regime, $k > 1/\eta$ become redefined as IR modes at later times when $k < 1/\eta_{\rm later}$. That is, in Eq.~\ref{eq:coincg}, we should consider $\Lambda \sim H$. Even when we turn interactions back on this diffusion effect will continue across the UV/IR boundary $k \sim H$, before the interactions become significant due to large logarithms. 
  
 Including interactions in the far IR is more subtle and interesting. To focus on it, we neglect the above diffusion effect of comoving modes redshifting from the UV to IR wavefunctional factors.
 At weak coupling we expect the full path integral evolution of $\Psi_{\rm IR}$ to be dominated by the stationary phase approximation, where the stationary $\phi$ solves the classical equations of motion, 
 \begin{equation}
 \partial_t^2 \phi + 3H \partial_t \phi  - e^{-2Ht} \nabla^2 \phi + V'(\phi) = 0.
 \end{equation}
 At late times the gradient terms are evidently unimportant due to redshifting, and the 
  the field acceleration becomes subdominant  to the friction term for weak coupling.
   Thus for 
 weak coupling and long time evolution (far IR), we expect the stationary phase field to satisfy
 \begin{equation}
 3H  \dot{\phi}(\vec{x}, t)  \approx - V'(\phi(\vec{x}, t)),
 \end{equation}
 which is ultralocal in space.  Furthermore, since it is first-order it is deterministic.  For an infinitesimal time-step, given $\phi_0(\vec{x})$ at time $t$, we have $\phi(\vec{x}) = \phi_0(\vec{x}) - V^\prime(\phi_0(\vec{x}))/3H\, dt$ at $t+dt$.  It leads to the following simple evolution of the probability distribution at the origin $\vec{x} = 0$,
\be
p(\phi, t+ dt) &=& \int d\phi_0 \, \delta(\phi - \phi_0 + V^\prime(\phi_0)/3H\,dt) \, p(\phi_0,t) \nn \\
&=& (1+V^{\prime\prime} (\phi)/3H \, dt ) p(\phi + V^\prime (\phi)/3H \, dt ,t) \nn \\ 
\Rightarrow \dot p(\phi, t) &=& \frac{1}{3H} \partial_\phi (V^\prime(\phi) p(\phi, t)).
\label{eq:driftonly}
\ee
This gives the interacting, or ``drift''-only portion of the Fokker-Planck equation (Eq.~\ref{eq:theheroenters}) we had derived in Section \ref{sec:fp}.  

Putting together the two effects on $p$-evolution discussed above, drift \& diffusion, then results in the full  Fokker-Planck equation. They are additive effects at the order $dt$ level needed to 
derive $\dot{p}$. This heuristic derivation is physically intuitive and attractive, but one can ask how systematically justified some of the key approximations are, such as the stationary phase approximation in the IR, the neglect of its field acceleration and gradient terms, and the free field approximation in the UV.  The answer is that our leading-log diagrammatic analysis {\it is} the systematic justification.

The heuristic derivation also gives us a physical picture of how to interpret the subleading diagrammatic corrections. Most obviously, the two-derivative corrections to the stationary phase equations of motion correspond to the two-derivative corrections to $G_R$ trees in the diagrammatic analysis, and can be treated systematically as higher-order perturbations.  
More interesting are the incorporation of interactions involving UV modes and quantum fluctuation corrections in the IR. We expect that these effects are captured by the subleading diagrams involving loops of $G_R$ (or $G_R$ lines that connect different correlator points), the regime of hard loop momenta corresponding to the UV interactions and soft loop momenta to IR quantum fluctuations. But we have not yet clearly established this.

We leave the operational formalism for including these corrections to future work.  In the parton shower of QCD, the Markovian picture is similarly spoiled at the NLL level.  Nonetheless, the subleading corrections take the form of isolated defects in the shower and remain subleading to the resummed LL contributions \cite{Baumgart:2010qf}.  Corrections to the parton shower can be implemented by a description using the density matrix \cite{Nagy:2007ty,Neill:2018uqw}.  It is suggestive that a similar framework may be of use here.\footnote{Ref.~\cite{Collins:2017haz} showed that time evolution of the density matrix in $\lambda \phi^4$, corrected perturbatively at fixed $\mo(\lambda)$ gives Fokker-Planck evolution for the diagonal entries.  It is certainly interesting that fixed-order calculations are sufficient for this result, but we have shown here why stochastic inflation is a correct description even after leading logs have necessitated working to arbitrary powers of $\lambda$.}

Beyond the theory of a light scalar on a fixed background, there are also generalizations of stochastic inflation that are both important and likely tractable.  As we mentioned in the Introduction, allowing the curvature to be dynamical and respond to the energy density in different Hubble patches given by $V(\phi)$ for varying $\phi$ effectively gives a landscape with an unambiguous measure function.  One could thus bring full QFT rigor to questions about eternal inflation and its possible viability as a resolution to the cosmological constant problem.  Another extension is to study low-energy quantum gravity itself on a DS background.  Some steps have been taken to understand possible infrared pathologies in DS theories involving gravity with or without matter \cite{Anninos:2014lwa,Rajaraman:2016nvv}, as well as consequences of backreaction on De Sitter geometry itself \cite{Geshnizjani:2003cn,Brandenberger:2018fdd}.  While graviton calculations are more technically challenging, there is nothing inherent to them that would spoil a similar analysis to ours.  In this work, we have adopted much of the terminology and insights of effective field theory.  Nonetheless, a complete Soft De Sitter Effective Theory (SDET), with a leading-power lagrangian, consistent operator power expansion, and RG resummation remains to be developed.

\vspace{0.3in}

{\it Acknowledgments:} 
We thank Nima Arkani-Hamed, Daniel Green, Juan Maldacena, Arvind Rajaraman, Ira Rothstein, and Leonardo Senatore for useful discussions.  MB is supported by the U.S. Department of Energy, under grant number DE-SC-0000232627. RS is supported by NSF grant PHY-1914731 and by the Maryland Center for Fundamental Physics (MCFP).

\appendix
\section{The Nested Commutator In-In Formalism}
\label{app:inin}

The standard, interaction-picture form of the in-in formalism is given by 
\be
\vev{ \phi_{\rm Heis.}(t,x)^n } &=& \bigg \langle \left[ \bar T \exp \left( i \int_{t_0}^t H_I(t^\prime) dt^\prime \right) \right] \nn \\
&&\times \phi_I(t,x)^n \left[ T \exp \left( -i \int_{t_0}^t H_I(t^{\prime\prime}) dt^{\prime\prime} \right) \right] 
\bigg \rangle .
\label{eq:appintev}
\ee
For our purposes, it is more convenient to use the nested commutator form, found in \cite{Weinberg:2005vy},
\be
\vev{\phi_{\rm Heis.}(t,x)^n} &=& \sum_{V=0}^\infty (-i)^V \int_{t_0}^t dt_V \ldots \int_{t_0}^{t_3} dt_2 \int_{t_0}^{t_2} dt_1 \nn \\
&\times & \Big \langle \Big [ \Big [ \ldots \Big [ \phi_I(t,x)^n,  H_I(t_V) \Big ] \ldots , H_I(t_2) \Big ], H_I(t_1) \Big ] \Big \rangle \nn \\
&\equiv & \sum_{V=0}^\infty \vev{\phi(t,x)^n}\big |_{\lambda^V}.
\label{eq:apppertex}
\ee
While \cite{Weinberg:2005vy} sketches a proof of the equality of these two equations, we provide a full one here.

It is straightforward at zeroth and first order in the interaction.  We proceed by induction, assuming it to hold at $(V-1)^{\rm th}$ order.  Next, we take a time derivative of a modified version of the expression in Eq.~\ref{eq:appintev},
\beq
\frac{d}{d\tilde{t}} \Big \langle U^\dag(\tilde t,\,t_0) \, \phi_I(t,x)^n \, U(\tilde t,\,t_0) \Big \rangle
= \; -i\, \langle U^\dag(\tilde t,\,t_0) \, \left[\phi_I(t,x)^n, H_I(\tilde t) \right] \, U(\tilde t,\,t_0) \Big \rangle,
\eeq
where $U(t_2,\,t_1) \equiv T \exp \left( -i \int_{t_1}^{t_2} H_I(t^{\prime}) dt^{\prime} \right)$.
However, we have shown that expectation value of $\left[ \phi_I(t,x)^n , H_I(\tilde t) \right]$ is the same if we expand the $U^\dag,\,U$ operators to $(V-1)^{\rm th}$ order in Eq.~\ref{eq:appintev} or use $(V-1)$ nested commutators in Eq.~\ref{eq:apppertex}.  Thus, we can rewrite it as
\be
\vev{\left[\phi_I(t,x)^n, H_I(\tilde t) \right]}\big |_{V-1} &=& (-i)^{V-1} \int_{t_0}^{\tilde t} dt_{V\!-\!1} \ldots \int_{t_0}^{t_2} dt_1 \nn \\
&\times & \Big \langle \Big [ \ldots \Big [ \Big[ \phi_I(t,x)^n,  H_I(\tilde t) \Big ], H_I(t_{V\!-\!1}) \Big] \ldots , H_I(t_1) \Big ] \Big \rangle,
\ee
but this is just the derivative with respect to $\tilde t$ of Eq.~\ref{eq:apppertex} at $V^{\rm th}$ order if the upper limit of the $t_V$ integral is changed from $t$ to $\tilde t$.  Thus, we have that the $\tilde t$ derivatives of the two expressions are equal up to $V^{\rm th}$ order, and since the expressions themselves are equal at all orders for $\tilde t \rightarrow -\infty$, then they are also equal for general $\tilde t$ at $V^{\rm th}$ order.  Setting $\tilde t = t$ shows the equality of Eqs.~\ref{eq:appintev} and \ref{eq:apppertex}.

\section{Closing Loopholes in the Leading-Log Argument}
\label{app:closel}

In Section \ref{subsec:llexp}, we derived that to all orders in perturbation theory, a diagram with $P$ propagators has $\log(\kir \eta)^P$ leading dependence on the infrared cutoff.  We gave a simple power-counting argument, combined with causality, that the diagrams we considered, with only $V$ retarded propagators, and the approximations we took (expanding in soft $k$ and strongly-ordered times, $|\eta_{\rm early}| \gg |\eta_{\rm late}|$) give the maximal sensitivity to $\kir$.  While the argument is highly plausible and physical (and ultimately correct), it does have logical loopholes. Here we complete the proof by demonstrating the subleading nature of taking additional $G_R$ propagators or moving away from the soft, strongly-ordered regime. 
 
It is straightforward to consider the case of diagrams with a general number of $G_R$ propagators (but still $\geq V$ in order to get a nonzero result by satisfying the causality constraints).  We begin by continuing to use strong-ordering in time, $(|\eta_{\rm early}| \gg |\eta_{\rm late}|)$, as well as the leading soft limits of $G_R, G_+$, 
 and later show that these approximations too can be relaxed to full generality.   We have
\be
\vev{\phi(\eta,0)^n} \Big |_{\lambda^V} &\sim& \lambda^V \, \int_{1/\kir}^{A\, \eta} \frac{d\eta^{(V)}}{(H \eta^{(V)})^4} \ldots
\int_{1/\kir}^{A\, \eta^{(2)}} \frac{d\eta^{(1)}}{(H \eta^{(1)})^4} \nn \\
&&\times \int_{\kir}^{1/\eta_{\rm earliest, 1}} \frac{d^3k_1}{(2\pi)^3} \ldots \int_{\kir}^{1/\eta_{\rm earliest, \pmv}} \frac{d^3k_{\pmv}}{(2\pi)^3} \nn \\
&&\times \prod_{i=1}^{N_R} G_{R\, {\rm soft}} \left(\eta^{(i_1)},\, \eta^{(i_2)} \right) \, \prod_{j=1}^{P-N_R} G_{+\, {\rm soft}} \left(k_j \right). \nn \\
&&+\; ({\rm Non\textnormal{-} strongly\textnormal{-} ordered\ contributions}).
\label{eq:mixedcase}
\ee
The strong-ordering in time is defined and imposed in terms of a modestly big constant, $A$, as already introduced below Eq.~\ref{eq:effgrexpl}. 
We have put in the effective UV cutoffs on the momentum integrals beyond which there are rapidly oscillating phases of the full Green's functions ({\it cf.}~Eq.~\ref{eq:pmprops}).  In the strongly-ordered limit, each momentum integral is cut off beyond $k_j \sim 1/\eta_{\rm earliest,\, j}$, the inverse of the earliest conformal time of the vertex that momentum flows through.  For the integrals in Eq.~\ref{eq:mixedcase}, each time, $\eta^{(i)}$, is the earliest time for some number of retarded propagators, $N_i$.\footnote{One may recall that in the leading-log case with only $V$ retarded propagators, each $\eta^{(i)}$ is the earliest time for exactly one $G_R$.}  Thus, it contributes $\eta^{(i) \, 3N_i}$ to the integrand. 
We can again assign momenta such that only a single loop momentum flows through each $G_+$.  These observations then give
\be
\vev{\phi(\eta,0)^n} \Big |_{\lambda^V} &\sim& \lambda^V \, \int_{1/\kir}^{A\, \eta} \frac{d\eta^{(V)}}{(H \eta^{(V)})^4} \ldots
\int_{1/\kir}^{A\, \eta^{(2)}} \frac{d\eta^{(1)}}{(H \eta^{(1)})^4} \nn \\
&&\times \int_{\kir}^{1/\eta_{\rm earliest, 1}} \frac{d^3k_1}{(2\pi)^3} \ldots \int_{\kir}^{1/\eta_{\rm earliest, \pmv}} \frac{d^3k_{\pmv}}{(2\pi)^3} \prod_{i=1}^{N_R} \eta^{(i)\, 3 N_i} \, \prod_{j=1}^{P-N_R} k_j^{-3}, 
\label{eq:mixedcaseexpl}
\ee
when we plug in the soft limits of the $G_R, G_+$ propagators. Thus, as in the main text, for every $k_j$ that appears in a $G_+$, its integral will give $\log(\kir \, \eta_{\rm earliest, j})$.

\begin{figure}
\begin{center}
\includegraphics[width=12cm]{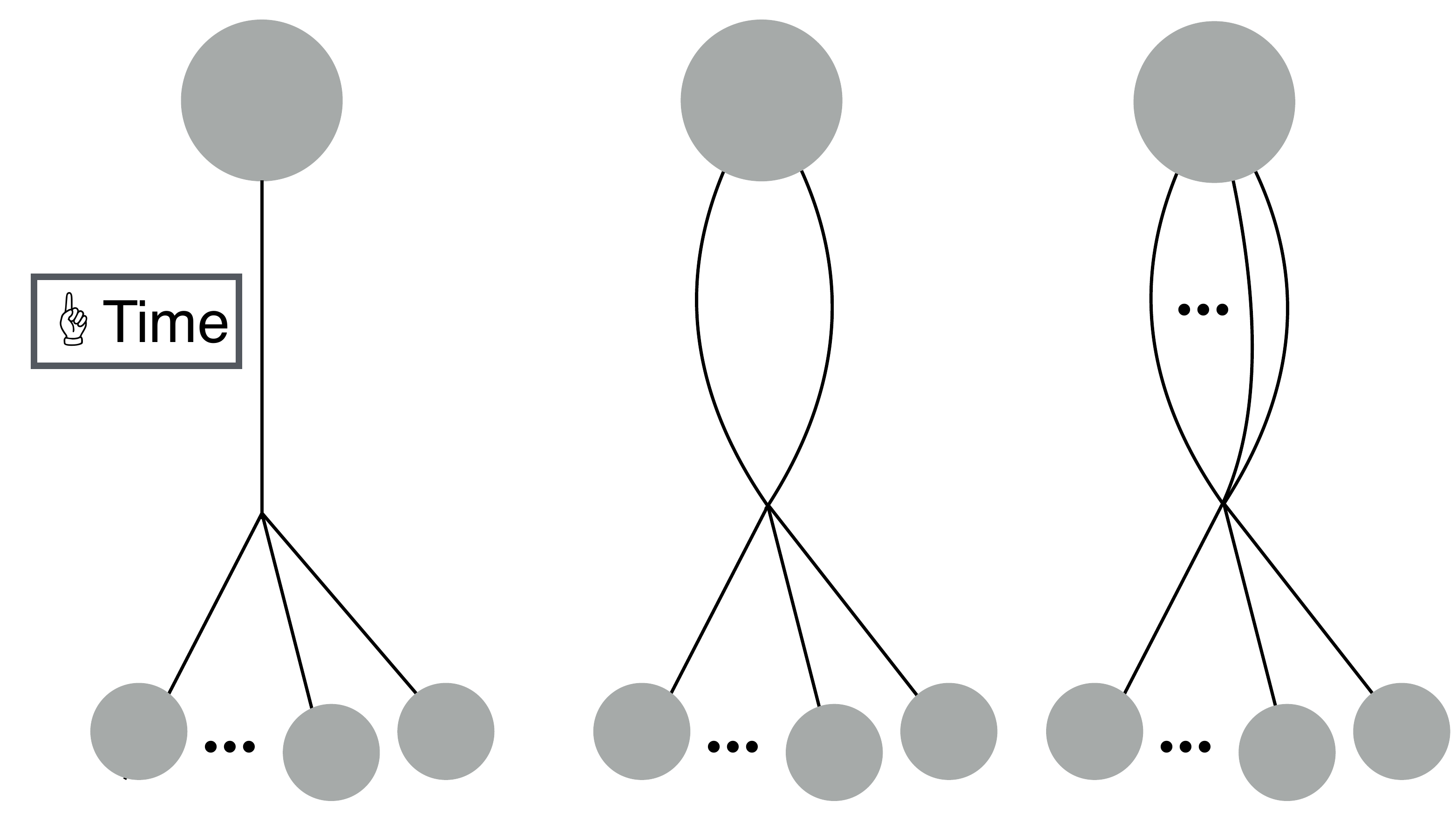}    
\end{center}
\caption{These diagrams show different numbers of $G_R$ entering a general vertex, at time $\eta^{(i)}$, from the future, where all $G_+$ lines are omitted.  This illustrates for how many loop momenta that vertex is the earliest time they touch.  In the left case, with only one retarded propagator from the future, its momentum necessarily flows further into the past, and $\eta^{(i)}$ is the earliest time for zero momenta.  In the middle case, having two retarded propagators enter the vertex means a loop momentum can be routed down one and up the other, making $\eta^{(i)}$ the earliest vertex through which that momentum flows.  In the general case, shown on the right, the argument generalizes such that if $\eta^{(i)}$ is the earliest time for $N_i$ retarded propagators, it is also the earliest time vertex that $N_i - 1$ loop momenta flow through.  This relation between the powers of $\eta^{(i)}$ appearing in $G_R$ factors in the integrand and the $\eta^{(i)}$ scaling of the loop momenta UV cutoffs determines the $\kir$ scaling of the graph given in Eq.~\ref{eq:genscaling} for the generic arrangement of $G_R$ and $G_+$ propagators.   
\vspace{0.025in}}
\label{fig:ev}
\end{figure}

Beyond the loop momenta associated to the $G_+$ lines, there are  in general {\it extra} loop momenta because the $G_R$-only subgraphs are no longer restricted to be trees as in the main text. The number of these extra loop momenta for which the $\eta^{(i)}$ vertex is the earliest they flow through is $N_i -1$.  This is straightforward to see iteratively ({\it cf.}~Fig.~\ref{fig:ev}).  
The associated loop integrals will each contribute a factor of $1/\eta^3_{\rm earliest,\, j}$, since all the $G_{R\, {\rm soft}}$ are $k$-independent. 
Thus, doing these loop integrals multiplies each factor of  $\eta^{(i)\, 3 N_i}$ in Eq.~\ref{eq:mixedcaseexpl} by $1/\eta^{(i)\, 3 (N_i - 1)}$, yielding a net $\eta^{(i)\, 3}$. 
Putting all the momentum integral results together, we have  
\be
\vev{\phi(\eta,0)^n} \Big |_{\lambda^V} &\sim& \lambda^V \, \prod_{i=1}^{V} \int_{1/\kir}^{A\, \eta^{(i+1)}} d\eta^{(i)} \eta^{(i)\, -1} \, 
\prod_{j=1}^{P-N_R} \log(\kir \, \eta_{\rm earliest,\, j}) ,  \nn \\
&\sim& \lambda^V \log(\kir \, \eta)^{V+P-N_R}. 
\label{eq:genscaling}
\ee
We have thus confirmed our expectation of the main text, that more $G_R$ propagators than minimally required by causality constraints, $N_R > V$, means a subleading dependence on $\kir$.  As we found in the leading-log case, the identity of the $\eta_{\rm earliest,\, j}$ in the $\log$ argument is ultimately irrelevant to the final result.

We can now move on to examine deviations from our strongly-ordered-in-time, and soft-momentum approximations.  Firstly, we claimed that the leading-log contribution occurs when times are strongly ordered, $|\eta^{(i+1)}| \ll |\eta^{(i)}|$.  Secondly, we massively simplified the UV behavior of our correlation functions by taking our propagators to be given by their leading behavior in small $k$, but then cut them off at the point where the full propagator becomes rapidly oscillating.  For a given loop momentum, its effective UV cutoff is given by the inverse of the time at the earliest vertex it flows through.  Since earlier times have larger absolute value, the trignometric factors in propagators ({\it e.g.}~$\sin \left[ k_m (\eta_{\rm early} - \eta_{\rm late}) \right]$) will be rapidly oscillating for $k_m \gtrsim 1/\eta_{\rm early}$.  Of course, there is a regime where $\eta_{\rm early} \sim \eta_{\rm late}$, in which case the regime of rapid oscillation is much larger, $k_m \gtrsim 1/|\eta_{\rm early} - \eta_{\rm late}|$.  But even in this near-coincident time regime, the comoving momenta will be cut off by the ultimate physical cutoff 
$\Lambda/(H\, \eta)$.\footnote{Reference to $\Lambda$ can eventually be eliminated by renormalization in favor of a renormalization scale, but this does not affect the simple point we make here.} For uniform treatment of pairs of times that are strongly ordered and times that are nearly coincident, we can simply consider this maximal momentum cutoff  $\Lambda/(H \eta_{\rm early})$, since even if the times are strongly ordered, rapid phase cancelation will kick in anyway for momenta $\gtrsim 1/\eta_{\rm early} < \Lambda/(H \eta_{\rm early})$. 

We see that properly including the regimes of near-coincident times into Eq.~\ref{eq:phischem} can be accomplished by replacing the momentum integral cutoffs by $\Lambda/(H \eta_{\rm early})$ rather than just $1/\eta_{\rm early}$, and by 
keeping the full $\eta$-dependence in $G_{R \, {\rm soft}}$ of $(\eta^{(i+1)\, 3} - \eta^{(i)\, 3})$ replacing $\eta^{(i)\, 3}$.  For the purpose of { \it bounding} contributions that violate strong-ordering though, we note we can just retain $\eta^{(i)\, 3}$ as an upper bound on $(\eta^{(i+1)\, 3} - \eta^{(i)\, 3})$. Further, note the $\Lambda/H$ factors in the momentum cutoff cannot in themselves change the counting of log$(k_{\rm IR} \eta)$ powers that we obtained assuming strongly-ordered conformal times. 
Unlike the strongly-ordered regime, the $\eta^{(i)}$ integral in the region where $\eta^{(i)} \sim \eta^{(i+1)}$ is $\kir$-independent.  Since the $1/\kir$ lower limit of  conformal time integration cuts off a divergence, the correction to strong-ordering that occurs for each $\eta$ integral in the $\eta^{(i)} \sim \eta^{(i+1)}$ region is necessarily subleading in $\kir$.  

We now turn to the effective UV cutoff.  We know that for small $k$ we have a consistent expansion.  We also know that the rapidly oscillating behavior will cause the integral to have vanishing support at very large $k$.  The issue is whether the ``missing'' powers of $k$, those that appear neither in the trigonometric functions at large $k$, nor in the soft $k$ expansion, can make a parametrically significant contribution.  For example, in the case where $k \eta$ is large enough to make a significant perturbative correction, but $k \eta \lesssim 1$ so that we do not get rapid oscillation, we get a correction in $G_+$ like 
\beq
G_+ \sim \frac{H^2}{k^3} (1 + k^2 \eta_{\rm earliest, j}^2).  
\label{eq:etacrxn}
\eeq
Upon $k$ integration, the correction term will contribute an extra factor of $(1/\eta_{\rm earliest, j}^2)$  relative to the leading term, but we see this merely cancels the explicit $\eta_{\rm earliest, j}^2$ in the correction term.  It also lacks the logarithmic divergence arising in the leading term, and therefore gives only a subleading contribution in $k_{\rm IR}$. If we replace powers of $\eta_{\rm earliest, j}$ with $\eta_{\rm latest, j}$, then our correction would be further suppressed by some power of $|\eta_{\rm latest, j}/\eta_{\rm earliest, j}| < 1$, and would therefore remain subleading. The same points are true of the trignometric factors in the non-oscillating regime, $k \eta \lesssim 1$, where they can be Taylor expanded. Parallel statements hold for $G_R$.  

In conclusion all, departures from strong-ordering and from the soft approximations to the propagators are subleading in $k_{\rm IR}$.

\section{Restoring Acceleration}
\label{app:accel}

\begin{figure}
\begin{center}
\includegraphics[width=15cm]{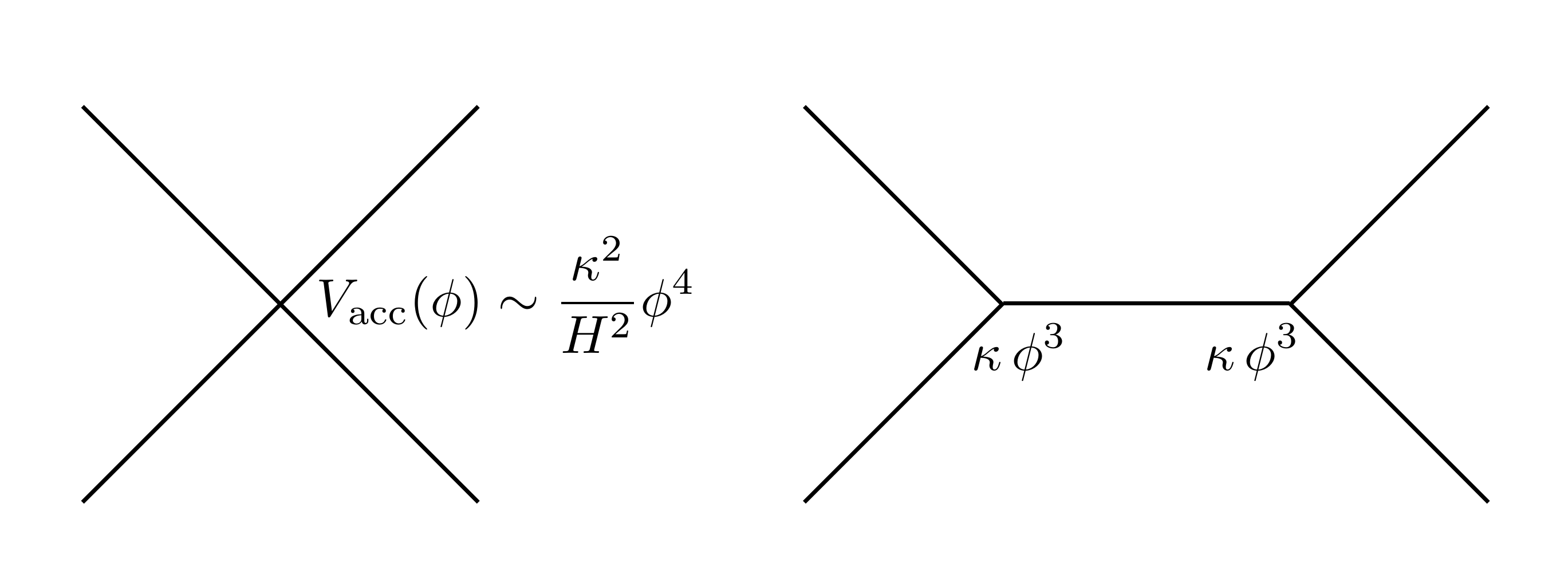}
\caption{We can include the effect of the $\ddot \phi$ term in the equation of motion by an effective vertex, $V_{\rm acc}(\phi)$, given in Eq.~\ref{eq:vacc}.  However, any diagram containing this vertex will be subleading to one using two insertions of the lagrangian interaction connected by a propagator, since the late time scaling of any graph is $t^P$ ({\it cf.}~Eq.~\ref{eq:llscaling}), where $P$ is the total number of propagators.  We show this explicitly here for contributions to $\vev{\phi^4}$ in $\phi^3$ theory.}
\label{fig:accvtx}
\end{center}
\end{figure}

We can ask if the acceleration term can ever return to importance (we have no such worry for the gradient term as its smallness defines the superhorizon regime, and it becomes monotonically smaller as time gets later).  Having established that the leading approximation involves just the first order equation of motion, we can add in the acceleration as a perturbation and check if it remains small.  We have 
\be
\ddot \phi \equiv \partial_t (\dot \phi) &=& \partial_t \left[ -\frac{1}{3H} V^\prime(\phi) \right] \nn \\
&=& \frac{1}{9H^2} V^\prime(\phi) V^{\prime \prime}(\phi).
\ee
Thus, if $V(\phi)$ is polynomial potential $\sim \lambda \phi^m$, then the acceleration generates an effective interaction,
\beq
V_{\rm acc}(\phi) \sim \lambda^2 \phi^{2m-2}.
\label{eq:vacc}
\eeq
As we have shown in Section \ref{sec:llcorr}, for any diagrams with a given topology, after appropriately summing over different in-in configurations, they give a contribution to an $n$-point correlation function, $t^P$, where $P$ is the number of propagators in the diagram.  Thus, any diagram with an insertion of the $V_{\rm acc}$ vertex will contribute $2m-2$ propagators and two powers of the perturbative coupling.\footnote{There is, of course, the possibility of contracting the $\phi$ fields from the vertex in Eq.~\ref{eq:vacc} with themselves, meaning that it adds fewer than $2n-2$ propagators to the diagram.  Nonetheless, it remains that case that one can replace the effective vertex with two $V(\phi)$ vertices connected by a propagator.  The argument is then the same.  This latter contribution still contributes more logs regardless of how many contractions one does among the $2m-2$ fields in either case.}  However this is included though, we will always have another contribution from two insertions of the original vertex in $V(\phi)$, connected by a single propagator (one gets back the effective vertex by contracting the line joining the two $V$ vertices to a point, {\it cf.}~Fig.~\ref{fig:accvtx}).  
This contribution with a pair of the original vertices involves the same number of powers of the coupling, two, and by construction it has an additional propagator, giving $2n-1$ in total.  Therefore, including the perturbation to the classical equation of motion due to the acceleration is a subleading-log effect.



\end{document}